\documentclass[12pt,a4paper]{article}
\usepackage{amsmath,amsfonts,xcolor}
\usepackage[pdftex,colorlinks=true]{hyperref}
\definecolor{darkblue}{rgb}{0,0,.6}
\hypersetup{citecolor=darkblue,linkcolor=darkblue,urlcolor=darkblue}
\usepackage[font={footnotesize,it}]{caption}
\usepackage[top=1.2in, bottom=1.2in, left=1.1in, right=1.1in]{geometry}
\usepackage[title]{appendix} 

\DeclareCaptionStyle{italic}[justification=centering]{labelfont={bf},textfont={it},labelsep=colon}
\usepackage{graphicx,psfrag,epsf}
\usepackage{enumerate}
\usepackage{natbib, orcidlink}
\usepackage{url, setspace}
\usepackage{booktabs, subfig, bm, paralist,mathpazo,tikz,todonotes,longtable,microtype,dsfont,rotating}

\newcommand{\blind}{0}

\newcommand{\Rlogo}{\protect\includegraphics[height=1.8ex,keepaspectratio]{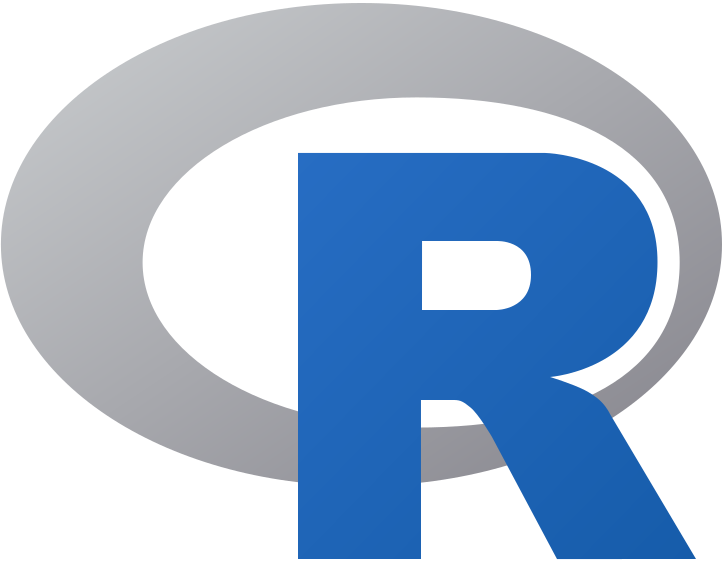}}

\newcommand{\E}{\text{E}}

\graphicspath{{plots/}}

\newsavebox\CBox
\def\textBF#1{\sbox\CBox{#1}\resizebox{\wd\CBox}{\ht\CBox}{\textbf{#1}}}

\date{}

\begin{document}

\def\spacingset#1{\renewcommand{\baselinestretch}%
{#1}\small\normalsize} \spacingset{1}

\if0\blind
{
  \title{\bf Weighted compositional functional data analysis for modeling and forecasting life-table death counts}
  \author{\normalsize Han Lin Shang \orcidlink{0000-0003-1769-6430} \footnote{Corresponding author. Telephone number: +61(2) 9850 4689; Email: hanlin.shang@mq.edu.au}
  \hspace{.2cm}\\
\normalsize     Department of Actuarial Studies and Business Analytics \\
\normalsize     Macquarie University \\
\\
\normalsize     Steven Haberman  \orcidlink{0000-0003-2269-9759} \\
\normalsize     Bayes Business School \\
\normalsize     City St George's, University of London
}
  \maketitle
} \fi

\if1\blind
{
    \title{\bf Weighted compositional data analysis for modeling and forecasting life-table death counts}
  \maketitle
} \fi

\begin{abstract}
Age-specific life-table death counts observed over time are examples of densities. Non-negativity and summability are constraints that sometimes require modifications of standard linear statistical methods. The centered log-ratio transformation presents a mapping from a constrained to a less constrained space. With a time series of densities, forecasts are more relevant to the recent data than the data from the distant past. We introduce a weighted compositional functional data analysis for modeling and forecasting life-table death counts. Our extension assigns higher weights to more recent data and provides a modeling scheme easily adapted for constraints. We illustrate our method using age-specific Swedish life-table death counts from 1751 to 2020. Compared to their unweighted counterparts, the weighted compositional data analytic method improves short-term point and interval forecast accuracies. The improved forecast accuracy could help actuaries improve the pricing of annuities and setting of reserves.

\vspace{.1in}
\end{abstract}

\newpage
\spacingset{1.5}

\section{Introduction}

Actuaries and demographers have long been interested in developing mortality modeling and forecasting methods. In the literature on human mortality, three functions are widely considered: hazard, survival, and probability density functions. Although these functions are complementary \citep{PHG01, DHW09}, most attention was given to new approaches for forecasting age-specific hazard function \citep[see, e.g.,][for reviews]{Booth06, BT08}. Instead of modeling central mortality rates, we consider modeling the life-table death distribution \citep[see, e.g.,][]{BKC20}. Observed over a period, we could model and forecast a redistribution of the density of life-table death counts, where deaths at younger ages are shifted gradually toward older ages. The period life-table death counts represent the mortality conditions, which, in recent years, have reflected a trend toward increasing longevity. Apart from providing an informative description of the mortality experience of a population, the life-table death counts yield readily available information on `central longevity indicators' \citep[see, e.g.,][]{CRT+05, Canudas-Romo10}, and lifespan variability \citep[see, e.g.,][]{Robine01, VZV11, HOC+13, VC13, VMM14, AV18, AVB+20}. 

In demography, \cite{Oeppen08} and \cite{BC17, BSO+18} treat life-table death counts as compositional data and use compositional data analysis (CoDa) to model and forecast age distribution of death counts. The data are constrained to vary between two limits (e.g., 0 and a constant upper bound), which in turn imposes constraints upon the variance-covariance structure of the original data. To remove the non-negativity and most of the summability constraints, the centered log-ratio transformation \citep{AS80, Aitchison82, Aitchison86} can be deployed before applying principal component analysis to the transformed data.

In actuarial science, \cite{SH20} and \cite{SHX22} apply the centered log-ratio transformation within the CoDa framework to model and forecast life-table death counts. The main contribution of this paper is that we extend the CoDa by assigning a set of geometrically decaying weights to estimate the geometric mean function and the estimated principal component decomposition. As described in Section~\ref{sec:3}, our extension assigns higher weights to relatively more recent data to improve short-term forecast accuracy. This extension is particularly important in demography, where we can have over 250 years of data, and data from the 18\textsuperscript{th} and 19\textsuperscript{th} centuries may not be so helpful in determining the recent trend for forecasting. 

In statistics, \cite{SDG+17} apply CoDa to study the concentration of chemical elements in sediment or rock samples. \cite{SW17} apply CoDa to analyze total weekly expenditure on food and housing costs for households in a chosen set of domains. \cite{SM22} apply CoDa to model cause-specific mortality data. \cite{Delicado11} and \cite{RIM21} use CoDa to analyse density functions over space, while \cite{KMP+19} model and forecast a time series of density functions. 

Using Swedish life-table death counts from 1751 to 2020 in Section~\ref{sec:2}, we highlight the difference between the weighted and standard CoDa methods in Section~\ref{sec:3}. In Section~\ref{sec:4}, we study the optimal selection of weight parameters for each horizon by minimizing point and interval forecast errors. We evaluate and compare point forecast accuracy in Section~\ref{sec:5} and interval forecast accuracy in Section~\ref{sec:6}, respectively. The conclusion is presented in Section~\ref{sec:7}, along with some ideas on how the methodology can be further extended.

\section{Swedish age distribution of death counts}\label{sec:2}

\begin{figure}[!htb]
\centering
\includegraphics[width=8.2cm]{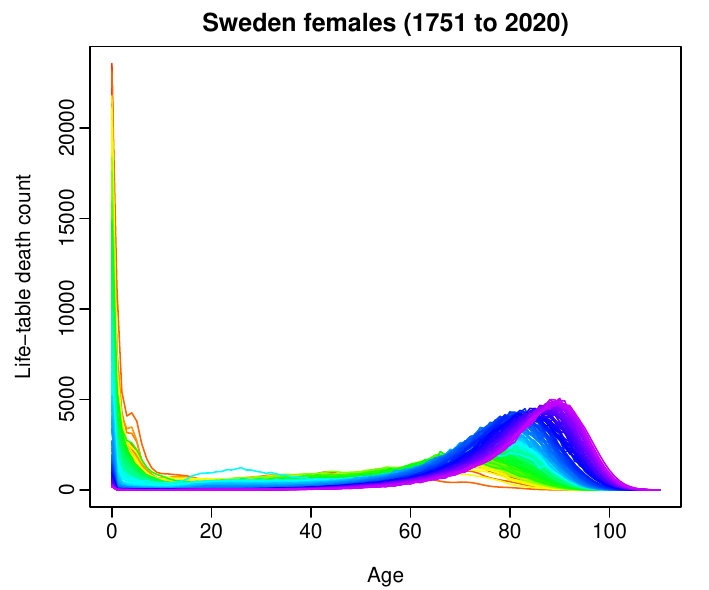} 
\qquad
\includegraphics[width=8.2cm]{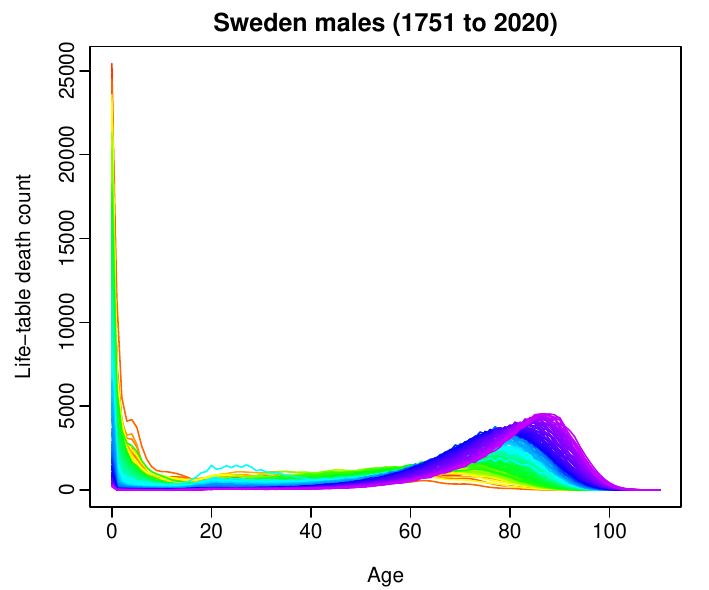} 
\caption{\small Rainbow plots of age-specific period life-table death count from 1751 to 2020 in a single-year group in Sweden. Curves are ordered chronologically according to the colors of the rainbow. The oldest years are displayed in red, with the most recent years shown in violet.}\label{fig:1}
\end{figure}

We consider age- and sex-specific life-table death counts from 1751 to 2020 in Sweden, the country with the longest record in the \cite{HMD22}. For over 250 years, each parish in Sweden has kept a complete and continuously updated register of its population \citep[see][]{GLW07}. We study life-table death counts, where the life-table radix (i.e., a population experiencing 100,000 births annually) is fixed at 100,000 at age 0 for each year. For the life-table death counts, there are 111 ages, and these are ages $0,1,\dots,109,110+$. Due to rounding, there are zero counts for age 110+ at some years, which may create an issue when taking log-ratio transformation. To rectify this problem, we use the probability of dying and the life-table radix to recalculate our estimated death counts (up to 6 decimal places). In doing so, we obtain more precise death counts than the ones reported in the \cite{HMD22}, which are often reported as integers. To some extent, the probability of dying relies on smooth rates \citep[see the][protocol for detail]{HMD22}. Figure~\ref{fig:1} presents rainbow plots of the female and male age-specific period life-table death counts in Sweden from 1751 to 2020 in single years.

Both sub-figures demonstrate a slowly decreasing trend in infant death counts and a typical negatively skewed distribution for the life-table death counts, where the peaks shift to higher ages for both females and males. This gradual shift is a primary driver of longevity risk, which is a major issue for insurers and pension funds, especially in the selling and risk management of annuity products \citep[see][for a discussion]{DDG07}.

The re-distribution of life-table death counts indicates lifespan variability across ages. A decrease in variability over time can be observed. This variability can be measured, for example, with the interquartile range of life-table death counts or the Gini coefficient \citep[for comprehensive reviews, see][]{WH99, SAB03, VC13, DCH+17}. In economics, the Gini coefficient summarizes the degree of concentration contained in the Lorenz curve with a single value, and its value varies from 0 (perfect equality) to 1 (perfect inequality). Since income and death counts are inversely related, a value of 0 indicates perfect inequality among ages in life-table death counts, and a value of 1 indicates perfect equality, implying that death occurs at the same age.
\begin{figure}[!htb]
\centering
\includegraphics[width=8.0cm]{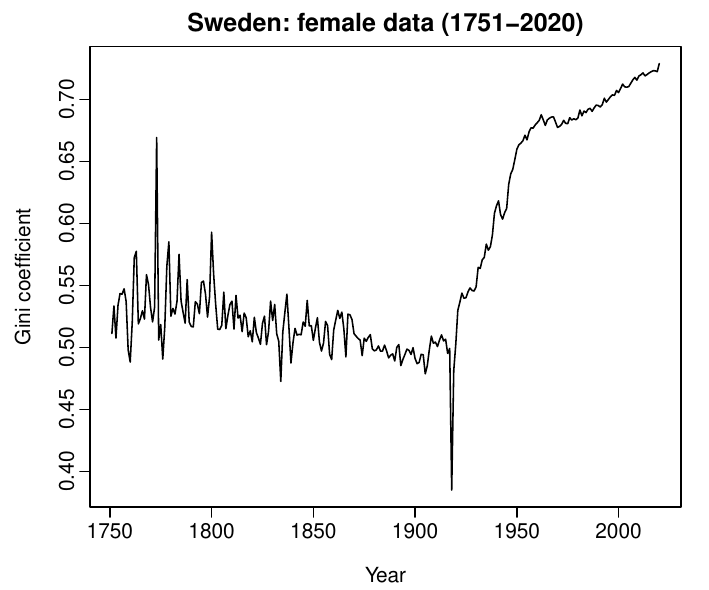}
\qquad
\includegraphics[width=8.0cm]{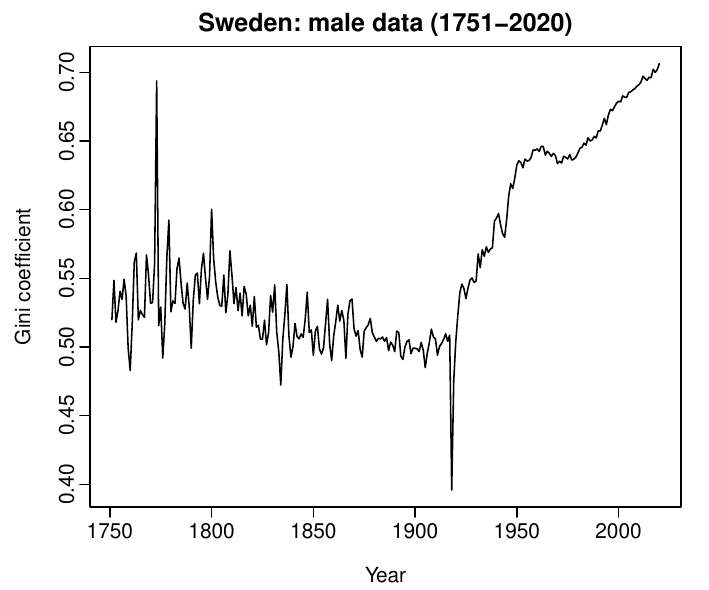}
\\
\includegraphics[width=8.0cm]{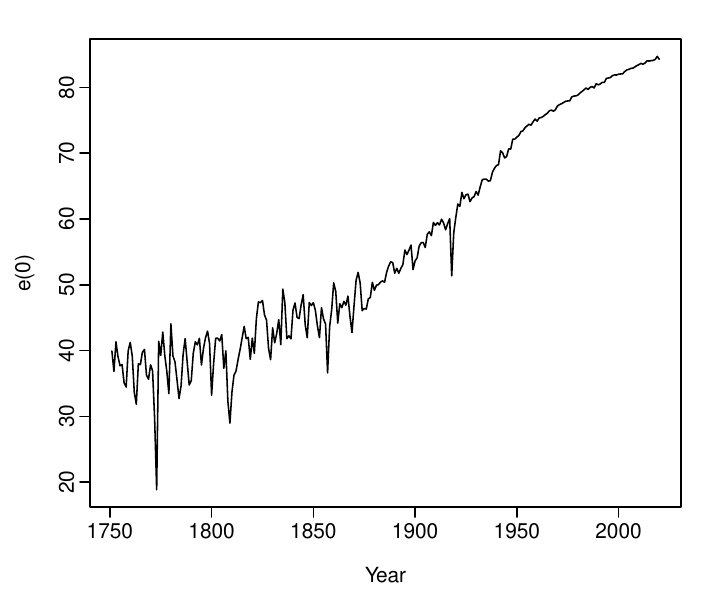}
\qquad
\includegraphics[width=8.0cm]{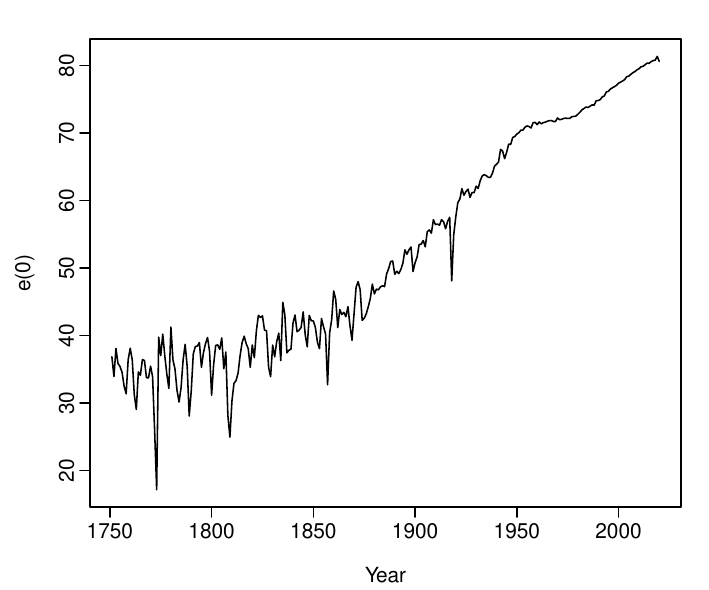}
\caption{\small Gini coefficients and life expectancy at age 0, e(0), for Swedish period female and male life-table death counts from 1751 to 2020. When the Gini coefficient approaches 0, it indicates perfect inequality across ages in the life-table death counts. When the Gini coefficient approaches 1, it indicates perfect equality.}\label{fig:2}
\end{figure}

Figure~\ref{fig:2} presents an example where the life-table death counts provide essential insights into longevity and lifespan variability that cannot be grasped directly from examining the trend in the age-specific central mortality rate or the survival function. We also include graphs of the trend in life expectancy at age zero, denoted by e(0), over time.

The Gini coefficient can be viewed as a single summary measure of inequality in a distribution. In economics, it is derived from the Lorenz curve which is non-decreasing. The Lorenz curve shares a strong resemblance to a cumulative distribution function. In economics, denote $\sf{p}$ as the fraction of the population that holds $L(\sf{p})$ proportion of the whole income. The Gini coefficient can then be expressed as
\[
G = 2\int^1_0 [\sf{p}-L(\sf{p})]d\sf{p}.
\]

The distribution of death counts has an inverse relationship to the income. When the coefficient equals 0, it expresses maximal age-at-death inequality. Inversely, when the coefficient equals 1, equality occurs for all ages. From Figure~\ref{fig:2}, the effects of the cholera epidemic that occurred in 1834 are apparent for the Swedish female and male data \citep{Larsson20}. In 1918, there was a sudden drop in the Gini coefficient related to the Spanish flu pandemic.

\section{Geometrically weighted compositional data analytic approach}\label{sec:3}

Density functions are non-negative functions that integrate into one. They share features with compositional data \citep[see, e.g.,][]{Aitchison86, PET15}. Compositional data arise in many scientific fields, such as geology (geochemical elements), economics (income/expenditure distribution), medicine (body composition), the food industry (food composition), chemistry (chemical composition), agriculture (nutrient balance bionomics), environmental science (soil contamination), ecology (abundance of different species) and demography (life-table death counts). 

In statistics, \cite{SDG+17} uses CoDa to study the concentration of chemical elements in sediment. \cite{SW17} apply CoDa to analyze the total weekly household expenditure on food and housing costs. \cite{Delicado11}, \cite{KMP+19}, \cite{SH20}, and \cite{SHX22} use the centered log-ratio transformation to analyze density functions and implement principal component analysis on the unconstrained space. In demography, \cite{Oeppen08} and \cite{BC17} introduce a principal component analysis approach to forecast life-table death counts within the centered log-ratio transformation.

For a given year $t$, compositional data are defined as a random vector of $I$ non-negative components, $[d_t(u_1),\dots, d_t(u_I)]$, whose sum is a specified constant, such as one (portion), 100 (percentage), $10^5$ (radix) in life-table death counts, and $10^6$ (parts per million) in geochemical trace element compositions \citep[][p.1]{Aitchison86}. Between the non-negativity and summability constraints, the sample space of compositional data is a simplex:
\[
\mathcal{S}^I = \left\{[d_t(u_1),\dots,d_t(u_I)]^{\top}, \quad d_t(u_i)>0, \quad \sum^I_{i=1}d_t(u_i) = c\right\}, \qquad t=1,\dots,n,
\]
where $u$ denotes a continuum, such as age in this study, $c$ is a fixed constant, $^{\top}$ denotes vector transpose, and $\mathcal{S}$ denotes a simplex. The simplex sample space is a $I-1$ dimensional subset of the real-valued space $R^{I}$. The restriction of shares to the unit simplex sometimes leads to the lack of an interpretable covariance structure, which has been recognized by researchers in many fields \citep[see, e.g.,][]{Aitchison86, BPG96, FFM96}. 

The centered log-ratio transformation presents one of many possible ways to deal with the non-negativity constraint by transforming the raw data. The clr transformation \citep[see, e.g.,][]{AS80, Aitchison82, Aitchison86} is a mapping between the simplex to the hyperplane in the Euclidean space. PCA can be applied directly to this hyperplane. The algorithm for implementing the weighted CoDa method consists of the following steps:
\begin{enumerate}
\item[1)] Compute the geometric mean function with geometrically decaying weights. The mean function can be estimated by a weighted average
\begin{equation}
\alpha_n(u) = \exp\left\{\sum^n_{t=1}w_t \ln [d_t(u)]\right\},\label{eq:alpha}
\end{equation}
where $w_t =\kappa(1-\kappa)^{n-t}$ is a set of geometrically decaying weights with $0<\kappa<1$ and $\sum^n_{t=1}w_t=1$, $\ln(\cdot)$ denotes natural logarithm, and $d_t(u)>0$ denotes the age-specific life-table death count (see Section~\ref{sec:2} for the treatment of $d_t(u) = 0$). In Figure~\ref{fig:3}, we display geometrically decaying weights when the weight parameter $\kappa=0.05$ or 0.95, respectively. When $\kappa=0.05$, forecasts depend on more distant past observations. When $\kappa=0.95$, forecasts rely on the most recent observations.
\begin{figure}[!htb]
\centering
\includegraphics[width=10.5cm]{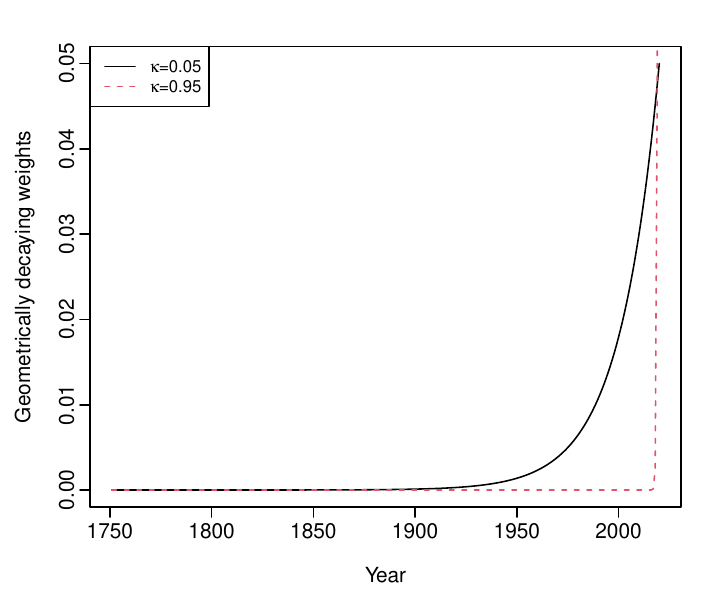}
\caption{\small Geometrically decaying weights when $\kappa=0.05$ and 0.95, respectively.}\label{fig:3}
\end{figure}

We treat age as a continuum $u\in [0,110]$ although age is observed at discrete points and set
\begin{equation}
s_t(u) = \frac{d_t(u)}{\alpha_n(u)}. \label{eq:st}
\end{equation}
\item[2)] Apply the centered log-ratio transformation given by
\begin{equation*}
\beta_t(u) = \ln[s_t(u)] \label{eq:log-ratio}
\end{equation*}
From~\eqref{eq:alpha} and~\eqref{eq:st}, we obtain
\begin{align}
\beta_t(u) &= \ln d_t(u) - \ln \alpha_n(u) \notag\\
&=\ln d_t(u) - \sum^n_{t=1}w_t\ln d_t(u)\notag
\end{align}
For a given population, $\beta_t(u)$ can be viewed as an unconstrained functional time series.
\item[3)] Compute eigenvalues and eigenfunctions from the covariance across $t$ of the weighted unconstrained functional time series: Let $\bm{\beta}(u) = [\beta_1(u),\dots,\beta_n(u)]^{\top}$ be a time series of unconstrained functions, and $\bm{w}=\text{diagonal}(w_1, \dots, w_n)$. The weighted curve time series is given as $\bm{\beta}^*(u) = \bm{w}\bm{\beta}(u)$. The weighting scheme aims to help find the most suitable basis functions for the curve time series to project onto. Applying eigendecomposition to $\bm{\beta}^*(u)$ gives 
\begin{equation}
\beta_t^*(u) = \sum^{n}_{k=1}\widehat{\gamma}_{t,k}\widehat{\phi}_k(u) = \sum^K_{k=1}\widehat{\gamma}_{t,k}\widehat{\phi}_k(u) + \omega_t(u),\label{eq:2}
\end{equation}
where $\widehat{\phi}_k(u)$ is the $k\textsuperscript{th}$ weighted orthonormal eigen-function (i.e., functional principal components), $\widehat{\gamma}_{t,k}=\langle \beta_t(u), \widehat{\phi}_k(u)\rangle$  is the $k\textsuperscript{th}$ set of principal component scores at time $t$, and $\omega_t(u)$ denotes the model residual function for age $u$ in year $t$.

When $K=1$,~\eqref{eq:2} reduces to the a version of the \citeauthor{LC92}'s \citeyearpar{LC92} method. The Lee-Carter model estimates parameters by minimizing the residual sum of squares via singular value decomposition. While the Lee-Carter model has the advantage of being nonparametric, its goodness-of-fit and forecasting performance can be improved with the inclusion of higher-order principal components \citep[see, e.g.,][]{BDV02, RH03, BT08}. We consider an eigenvalue ratio criterion of \cite{LRS20} to select the optimal value of $K$, and $K=6$ as used in \cite{HBY13} and the default opinion in the demography package \citep{Hyndman23}.
\item[4)] \textit{Forecasting $\bm{\beta}^*(u)$.} By conditioning on the estimated functional principal components $\bm{\Phi}(u) = [\widehat{\phi}_1(u),\dots,\widehat{\phi}_K(u)]$ and observed data $\bm{\beta}^*(u)$, the $h$-step-ahead forecast of $\bm{\beta}^*(u)$ can be obtained
\[
\widehat{\beta}_{n+h|n}^*(u) = \E[\beta_{n+h}(u)|\bm{\Phi}(u),\bm{\beta}^*(u)] = \sum^K_{k=1}\widehat{\gamma}_{n+h|n,k}\widehat{\phi}_k(u),
\]
where $\widehat{\gamma}_{n+h|n,k}$ denotes the $h$-step-ahead univariate time-series forecast of the $k$\textsuperscript{th} set of principal component scores. Among univariate time-series methods, we consider the random walk with drift (RWD) method for forecasting principal component scores.
\item[5)] \textit{Transform back to the compositional data.} We take the inverse centered log-ratio transformation given by
\[
\bm{\widehat{s}}_{n+h|n}(u)=\left[\exp^{\widehat{\beta}^*_{n+h|n}(u_1)},
\exp^{\widehat{\beta}^*_{n+h|n}(u_2)},\cdots,
\exp^{\widehat{\beta}^*_{n+h|n}(u_{111})}\right],
\]
where $\widehat{\beta}^*_{n+h|n}(u_i)$ denotes the forecasts in Step 3).
\item[6)] We add back the geometric means to obtain the life-table death count forecasts $\bm{d}_{n+h}(u)$,
\[
\widehat{\bm{d}}_{n+h|n}(u)=\left[\widehat{s}_{n+h|n}(u_1)\times \alpha_n(u_1),
\widehat{s}_{n+h|n}(u_2)\times \alpha_n(u_2),
\cdots,
\widehat{s}_{n+h|n}(u_{111})\times \alpha_n(u_{111})\right]
\]
where $\alpha_n(u_i)$ is the weighted geometric mean given in Step 1).
\end{enumerate}

To evaluate forecast uncertainty, we apply a nonparametric bootstrap method to generate future sample paths. We consider two sources of uncertainty: truncation errors in the functional principal component decomposition and forecast errors in the predicted principal component scores \citep[see, e.g.,][]{HS09, SH20}. Since principal component scores are regarded as surrogates of the original functional time series, these principal component scores capture the temporal dependence structure inherited in the original functional time series \citep[see, e.g.,][]{Paparoditis18, Shang18, PS22}. By adequately bootstrapping the forecast principal component scores, we can generate a set of bootstrapped forecasts $\bm{\widehat{\beta}}^*_{n+h|n}(u)=[\widehat{\beta}_{n+h|n}^{(1)}(u),\dots, \widehat{\beta}_{n+h|n}^{(B)}(u)]$, conditional on the estimated weighted mean function and weighted functional principal components from the observed data.

Using a univariate time series model, we can obtain multi-step-ahead forecasts for the principal component scores, 
$\{\widehat{\gamma}_{1,k},\dots,\widehat{\gamma}_{n,k}\}$ for $k=1,\dots,K$. Let the $h$-step-ahead forecast errors be
given by $\vartheta_{t,h,k}=\widehat{\gamma}_{t,k}-\widehat{\gamma}_{t|t-h,k}$ for $t=h+1,\dots,n$. The forecast errors, $\vartheta_{t,h,k}$, measure the difference between the estimated and forecast principal component scores. These can then be sampled
with replacement to give a bootstrap sample for $\gamma_{n+h,k}$:
\[
\widehat{\gamma}_{n+h|n,k}^{(b)} = \widehat{\gamma}_{n+h|n,k}+\vartheta^{(b)}_{*,h,k},\qquad b=1,\dots,B,
\]
where $B=1,000$ symbolizes the number of bootstrap samples and $\vartheta^{(b)}_{*,h,k}$ are sampled with replacement from $\{\vartheta_{h+1,h,k},\dots, \vartheta_{n,h,k}\}$.

Assuming the first $K$ functional principal components approximate the original functional time series relatively well, the model
residuals should contribute nothing but random noise. In contrast to the underestimation of $K$, the overestimation of $K$ does not hinder much forecast accuracy \citep[see also][]{HBY13}. Consequently, we can bootstrap the model fit errors in~\eqref{eq:2} by sampling with replacement from the model residual term $\{\omega_1(u),\dots,\omega_n(u)\}$.

Due to orthonormality, these two components of variability are summable. We obtain $B$ variants of $\beta_{n+h}(u)$,
\[
\widehat{\beta}_{n+h|n}^{(b)}(u) = \sum^K_{k=1}\widehat{\gamma}_{n+h|n,k}^{(b)}\widehat{\phi}_k(u)+\omega_{n+h}^{(b)}(u),
\]
where $\widehat{\gamma}_{n+h|n,k}^{(b)}$ denotes the bootstrap forecast of the principal component scores.

With the bootstrapped $\bm{\widehat{\beta}}^*_{n+h|n}(u)$, we follow Steps 4) and 5) of the above algorithm to obtain the bootstrap forecasts of $\bm{d}^{*}_{n+h|n}(u)$. At the $100(1-\alpha)\%$ nominal coverage probability, the pointwise prediction intervals are obtained by taking $\alpha/2$ and $1-\alpha/2$ quantiles based on $\{\widehat{d}_{n+h|n}^{(1)}(u),\dots, \widehat{d}_{n+h|n}^{(B)}(u)\}$.

\section{Selection of the geometrically decaying weight parameter}\label{sec:4}

\subsection{Point forecast error criteria}\label{sec:5.2}

Since the age-specific life-table death counts can be considered a probability density function, we also consider density evaluation metrics. They include discrete Kullback-Leibler divergence \citep{KL51} and the square root of the Jensen-Shannon divergence \citep{Shannon48, FT04}.

The Kullback-Leibler divergence measures information loss by replacing an unknown density with an approximation. For two probability density functions, denoted by $d_{n+\xi}(u)$ and $\widehat{d}_{n+\xi|n}(u)$, the discrete Kullback-Leibler divergence is defined as
\begin{align*}
\text{KLD}(h) = \ & D_{\text{KL}}[d_{n+\xi}(u_i)||\widehat{d}_{n+\xi|n}(u_i)]+D_{\text{KL}}[\widehat{d}_{n+\xi|n}(u_i)||d_{n+\xi}(u_i)] \\
= \ & \frac{1}{111\times (11-h)}\sum^{10}_{\xi=h}\sum^{111}_{i=1}d_{n+\xi}(u_i)\cdot [\ln d_{n+\xi}(u_i) - \ln \widehat{d}_{n+\xi|n}(u_i)] + \\
 & \frac{1}{111\times (11-h)}\sum^{10}_{\xi=h}\sum^{111}_{i=1}\widehat{d}_{n+\xi|n}(u_i)\cdot [\ln \widehat{d}_{n+\xi|n}(u_i) - \ln d_{n+\xi}(u_i)],
\end{align*}
where $i=111$ corresponds to the number of age groups, and $\xi$ corresponds to the forecasting period. The discrete Kullback-Leibler divergence is symmetric and non-negative.

An alternative is given by the Jensen-Shannon divergence, defined by
\[
\text{JSD}(h) = \frac{1}{2}\text{D}_{\text{KL}}[d_{n+\xi}(u_i)||\delta_{n+\xi}(u_i)] + \frac{1}{2}\text{D}_{\text{KL}}[\widehat{d}_{n+\xi|n}(u_i)||\delta_{n+\xi}(u_i)],
\]
where $\delta_{n+\xi}(u_i)$ measures a common quantity between $d_{n+\xi}(u_i)$ and $\widehat{d}_{n+\xi|n}(u_i)$. We consider the simple mean $\delta_{n+\xi}(u_i) = \frac{1}{2}[d_{n+\xi}(u_i) + \widehat{d}_{n+\xi|n}(u_i)]$ or geometric mean $\delta_{n+\xi}(u_i) = \sqrt{d_{n+\xi}(u_i)\widehat{d}_{n+\xi|n}(u_i)}$. We denote JSD$^{s}(h)$ for the Jensen-Shannon divergence with the simple mean and JSD$^g(h)$ for the Jensen-Shannon divergence with the geometric mean. 

\subsection{Interval forecast error criteria}\label{sec:5.3}

To evaluate the interval forecast accuracy, we consider the coverage probability difference (CPD) between the empirical and nominal coverage probabilities. For each year in the forecasting period, the $h$-step-ahead prediction intervals are computed at the $100(1-\nu)\%$ nominal coverage probability, where $\nu$ denotes a significance level, such as $\nu=0.2$ or 0.05. We consider the common case of the symmetric $100(1-\nu)\%$ prediction intervals, with lower and upper bounds that are predictive quantiles at $\nu/2$ and $1-\nu/2$, denoted by $\widehat{d}_{n+\xi}^{\text{lb}}(u_i)$ and $\widehat{d}_{n+\xi}^{\text{ub}}(u_i)$. The empirical coverage probability (ECP) is defined as
\[
\text{ECP}(h) = 1-\frac{1}{111\times (11-h)}\sum^{10}_{\xi=h}\sum^{111}_{i=1}\left[\mathds{1}\{d_{n+\xi}(u_i)>\widehat{d}_{n+\xi}^{\text{ub}}(u_i)\} + \mathds{1}\{d_{n+\xi}(u_i)<\widehat{d}_{n+\xi}^{\text{lb}}(u_i)\}\right].
\]
The CPD is then expressed as 
\[
\text{CPD}(h) = |\text{ECP}(h) - (1-\nu)|.
\]

\subsection{Expanding window scheme} \label{sec:5.3}

\begin{figure}[!htb]
\begin{center}
\begin{tikzpicture}
\draw (0,0) rectangle (15,2);
\draw (5,2) -- (5,0);
\draw (10,2) -- (10,0);
\draw (2.5,1) node {Training};
\draw (7.5,1) node[red] {Validation};
\draw (12.5,1) node {Testing};
\draw(2.5,2.2) node{1751:2000};
\draw(7.5,2.2) node{2001:2010};
\draw(12.5,2.2) node{2011:2020};
\end{tikzpicture}
\end{center}
\caption{\small Illustration of the cross-validation method. A model is constructed using data in the training set to forecast data in the validation set. The optimal weight parameter is determined based on the minimal forecast error in the validation set.}\label{fig:tikz}
\end{figure}
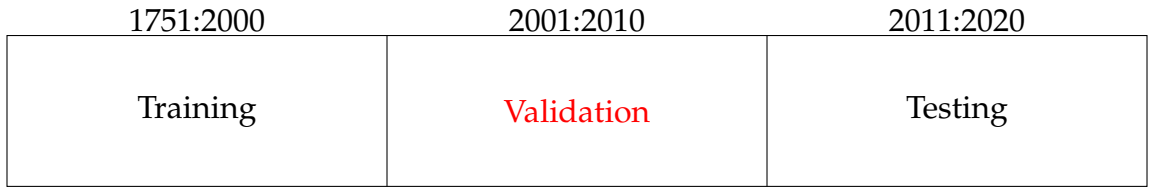

An expanding window scheme of a time series model is commonly used to assess model and parameter stability over time and the reliability of predictions. The expanding window analysis determines the variability of the model's parameters by computing parameter estimates and their forecasts over an expanding window. In Figure~\ref{fig:tikz}, we visually display sample splitting.

Using the first 250 observations from 1751 to 2000 in the life-table death counts, we produce one- to 10-step-ahead forecasts. Through an expanding window approach, we re-estimate the parameters using the first 251 observations from 1751 to 2001. Forecasts from the estimated models are then produced for one- to nine-step-ahead forecasts. We iterated this process by increasing the sample size by one year until we reached the data in 2010. This process produces 10 one-step-ahead forecasts, 9 two-step-ahead forecasts, $\dots$, and one 10-step-ahead forecast. We compare these forecasts with the validation samples from 2001 to 2010 to determine the optimal weight parameter, $\kappa$, for each of the ten forecast horizons. In Table~\ref{tab:1}, we tabulate the estimated geometrically decaying weight parameters in the weighted CoDa method under the KLD and two variants of the JSD.
\begin{table}[!htb]
\centering
\tabcolsep 0.32in
\caption{\small Estimated geometrically decaying weight parameters in the weighted CoDa method under the KLD and two variants of the JSD. The values in bold are used in our demonstration in Figure~\ref{fig:4}. The number of retained functional principal components is $K=6$ as used in \cite{HBY13}.}\label{tab:1}
\begin{tabular}{@{}lrrrrrrr@{}}
\toprule
	& \multicolumn{3}{c}{Female} 	& & \multicolumn{3}{c}{Male} \\\cmidrule{2-8}
$h$ & KLD & JSD$^s$ & JSD$^g$ & & KLD & JSD$^s$ & JSD$^g$ \\
\midrule
  1 & 0.024 & 0.024 & 0.024 & & 0.016 & 0.016 & 0.021 \\ 
  2 & 0.024 & 0.024 & 0.024 & & 0.019 & 0.029 & 0.029 \\ 
  3 & 0.049 & 0.049 & 0.025 & & 0.029 & 0.019 & 0.019 \\ 
  4 & 0.052 & 0.053 & 0.052 & & 0.076 & 0.076 & 0.076 \\ 
  5 & 0.055 & 0.055 & 0.055 & & 0.019 & 0.019 & 0.080 \\ 
  6 & 0.054 & 0.025 & 0.025 & & 0.030 & 0.030 & 0.030 \\ 
  7 & 0.056 & 0.056 & 0.056 & & 0.094 & 0.094 & 0.094 \\ 
  8 & 0.059 & 0.060 & 0.059 & & 0.093 & 0.093 & 0.093 \\ 
  9 & 0.064 & 0.065 & 0.065 & & 0.104 & 0.105 & 0.104 \\ 
  10 & 0.055 & 0.003 & 0.000 & & 0.106 & 0.106 & 0.106 \\ 
\bottomrule
\end{tabular}
\end{table}

\subsection{CoDa model fitting}

\begin{figure}[!htb]
\centering
\subfloat[Estimated geometric means]
{\includegraphics[width=8.2cm]{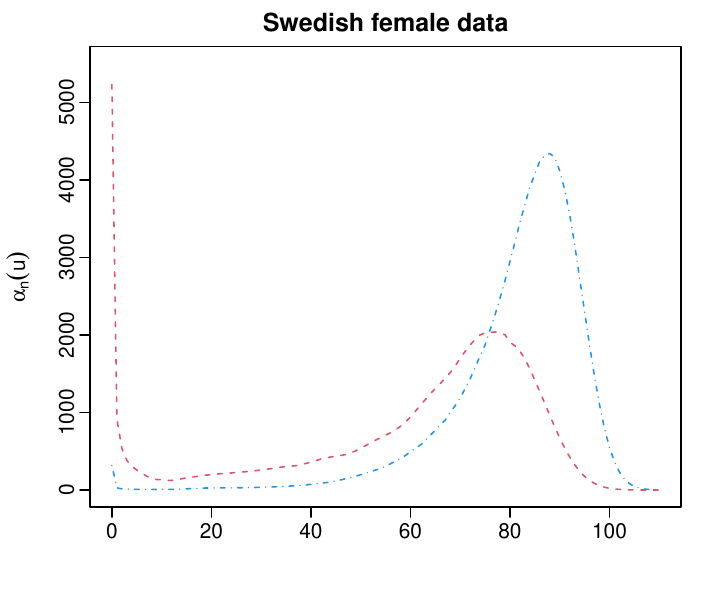}\label{fig:4a}}
\quad
\subfloat[Estimated geometric means]
{\includegraphics[width=8.2cm]{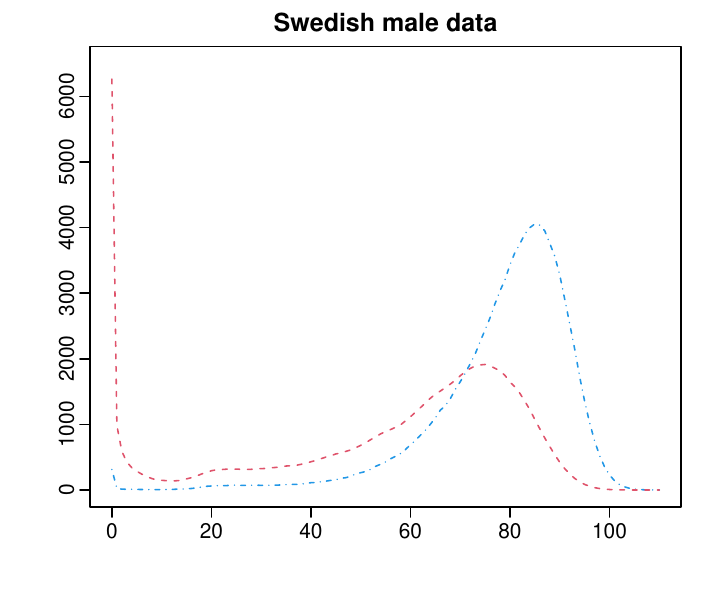}\label{fig:4b}}
\\
\subfloat[Estimated functional principal component]
{\includegraphics[width=8.2cm]{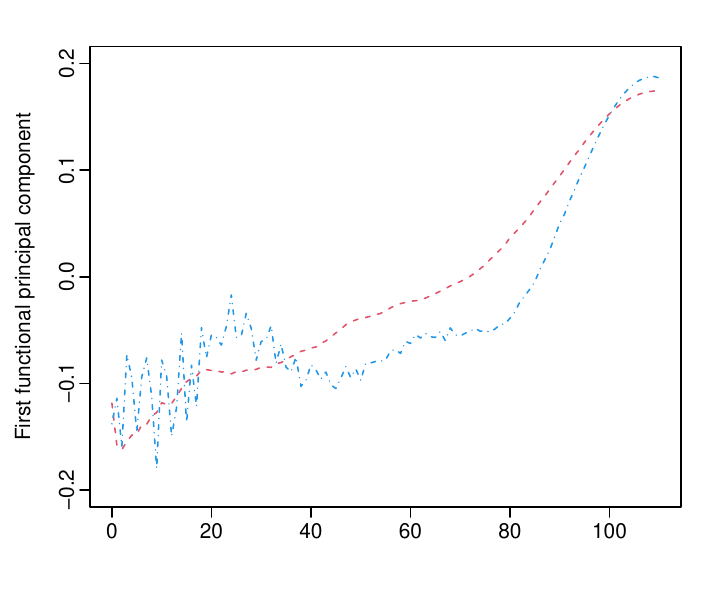}\label{fig:4e}}
\quad
\subfloat[Estimated functional principal component]
{\includegraphics[width=8.2cm]{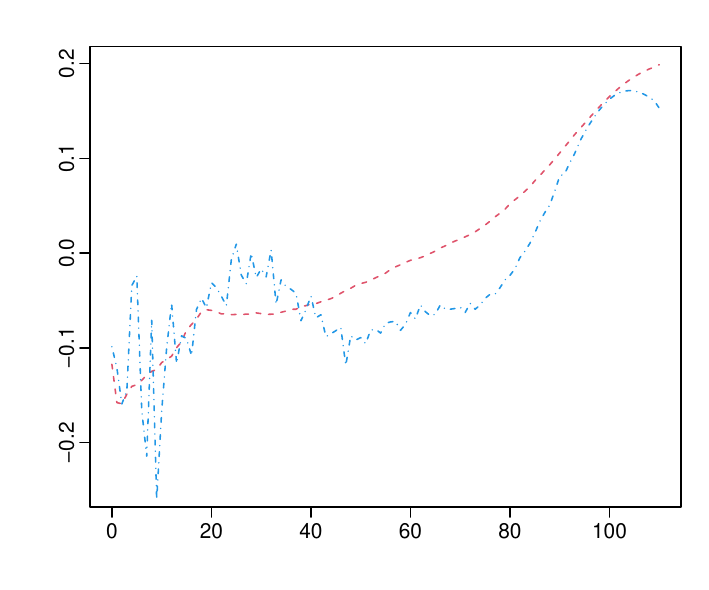}\label{fig:4f}}
\caption{\small Elements of the CoDa methods for modeling and forecasting the Swedish female and male life-table death counts. The estimated functions for the standard CoDa method are shown in red, while the estimated functions for the weighted CoDa method are displayed in blue.}\label{fig:4}
\end{figure}

We apply the standard and weighted CoDa methods to the Swedish female and male data. From the observed life-table death counts from 1751 to 2019 (i.e., 278 observations), we present the simple and weighted geometric means of the female and male life-table death counts in Figures~\ref{fig:4a} and~\ref{fig:4b}. The estimated functions for the standard CoDa method are shown in red, while those for the weighted CoDa method are displayed in blue. Via functional principal component analysis, we display the first estimated functional principal component in Figures~\ref{fig:4e} and~\ref{fig:4f}.

\begin{figure}[!htb]
\centering
\subfloat[One-step-ahead forecast life-table death counts]
{\includegraphics[width=8.2cm]{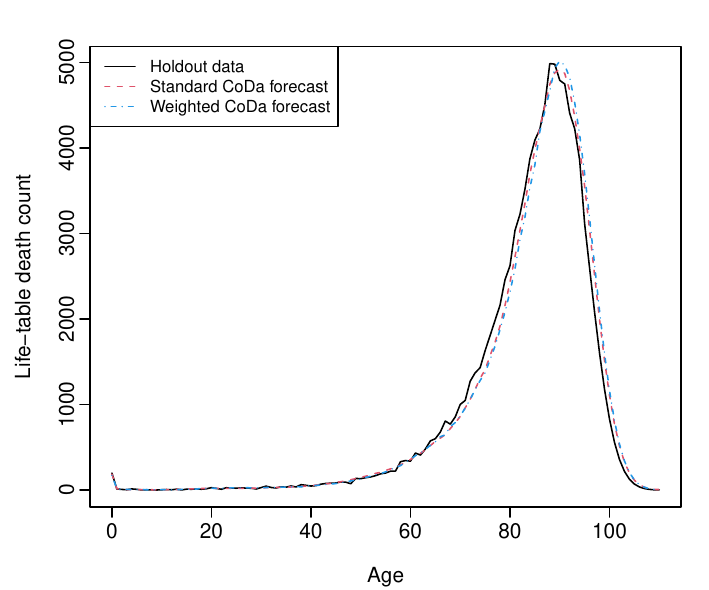}\label{fig:42g}}
\qquad
\subfloat[One-step-ahead forecast life-table death counts]
{\includegraphics[width=8.2cm]{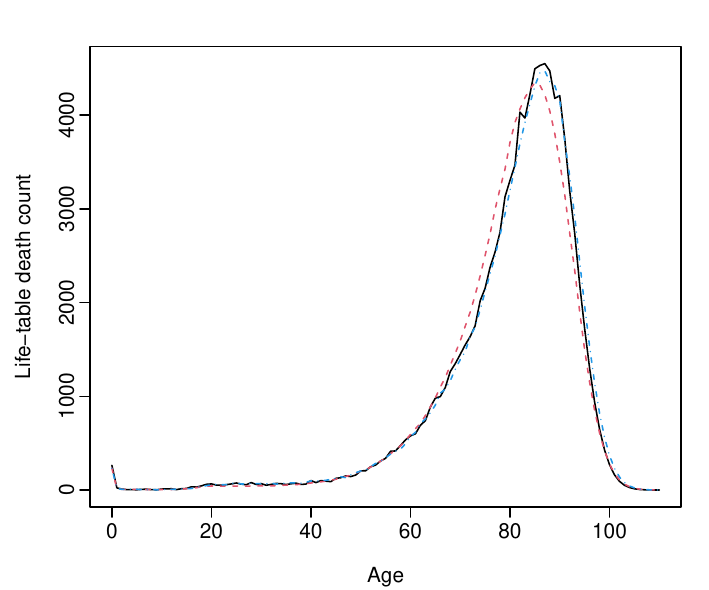}\label{fig:42h}}
\caption{\small One-step-ahead forecasts between the CoDa method shown in red and the weighted CoDa method in blue.}
\end{figure}

For producing one-step-ahead forecasts, the weight parameters $\kappa= 0.024$ and 0.016 are based on the KLD, as shown in Table~\ref{tab:1}. From the one-step-ahead point forecasts of life-table death counts in 2020, the weighted CoDa method produces one-step-ahead forecasts that are comparably closer to the holdout data as shown in Figures~\ref{fig:42g} and~\ref{fig:42h}.

\section{Comparison of point forecast accuracy}\label{sec:5}

Using the first 260 observations from 1751 to 2010, we produce one- to 10-step-ahead forecasts via an expanding window approach. We evaluate forecast accuracy by comparing the forecasts with the holdout data from 2011 to 2020. Table~\ref{tab:2} presents point forecast evaluation metrics between standard and geometrically decaying weighted CoDa methods for the Swedish female and male data. The weighted CoDa method's estimated weight parameters for different horizons are tabulated in Table~\ref{tab:1}. For the CoDa method, we consider using the fitting period from 1950 onwards and all available data.
\begin{singlespace}
\begin{center}
\tabcolsep 0.1in
\begin{longtable}{@{}lllrrrrrrrrr@{}}
\caption{\small Comparison of point forecast errors ($\times 100$), namely the KLD and two variants of the JSD, between the CoDa and its weighted variant for forecasting the age-specific Swedish female and male data. Based on the averages of the point forecast errors, minimum values are highlighted in bold. The eigenvalue ratio criterion is shorted as the EVR.} \label{tab:2} \\
\toprule
& & \multicolumn{3}{c}{CoDa} & \multicolumn{3}{c}{CoDa (1950)} & \multicolumn{3}{c}{Weighted CoDa} \\
Sex & $K$ & $h$ & KLD & JSD$^s$ & JSD$^g$ & KLD & JSD$^s$ & JSD$^g$ & KLD & JSD$^s$ & JSD$^g$ \\
\midrule
\endfirsthead
Sex & $K$ & $h$ & KLD & JSD$^s$ & JSD$^g$ & KLD & JSD$^s$ & JSD$^g$ & KLD & JSD$^s$ & JSD$^g$ \\
  \midrule
\endhead
\midrule \multicolumn{12}{r}{{Continued on next page}} \\
\endfoot
\bottomrule
\endlastfoot
Female 	& EVR & 1 & 1.278 & 0.314 & 0.315 & 0.207 & 0.049 & 0.049 & 0.185 & 0.043 & 0.043 \\ 
		& 	&  2 & 1.353 & 0.333 & 0.334 & 0.183 & 0.043 & 0.043 & 0.162 & 0.037 & 0.037 \\ 
		& 	&  3 & 1.448 & 0.356 & 0.358 & 0.199 & 0.046 & 0.047 & 0.171 & 0.039 & 0.040 \\ 
		& 	&  4 & 1.519 & 0.373 & 0.375 & 0.206 & 0.048 & 0.048 & 0.163 & 0.036 & 0.037 \\ 
		& 	&  5 & 1.519 & 0.373 & 0.375 & 0.247 & 0.058 & 0.058 & 0.206 & 0.047 & 0.048 \\ 
		& 	&  6 & 1.585 & 0.390 & 0.392 & 0.300 & 0.071 & 0.072 & 0.210 & 0.050 & 0.050 \\ 
		& 	&  7 & 1.683 & 0.414 & 0.417 & 0.357 & 0.086 & 0.086 & 0.226 & 0.053 & 0.053 \\ 
		& 	&  8 & 1.835 & 0.452 & 0.455 & 0.330 & 0.079 & 0.079 & 0.265 & 0.062 & 0.063 \\ 
		& 	&  9 & 1.886 & 0.464 & 0.468 & 0.431 & 0.104 & 0.105 & 0.299 & 0.070 & 0.071 \\ 
		& 	&  10 & 1.170 & 0.288 & 0.289 & 0.840 & 0.205 & 0.206 & 0.220 & 0.051 & 0.051 \\ 
\cmidrule{3-12}
		& 	& Mean 	& 1.527 & 0.376 & 0.378 & 0.330 & 0.079 & 0.079 & \textBF{0.211} & \textBF{0.049} & \textBF{0.049} \\ 
\cmidrule{2-12}
		& $K=6$ 	& 1 & 0.282 & 0.067 & 0.067 & 0.212 & 0.048 & 0.049 & 0.248 & 0.057 & 0.057 \\ 
		& 		& 2 & 0.309 & 0.073 & 0.074 & 0.198 & 0.044 & 0.045 & 0.232 & 0.052 & 0.052 \\ 
		& 		& 3 & 0.324 & 0.078 & 0.078 & 0.220 & 0.050 & 0.050 & 0.206 & 0.045 & 0.049 \\ 
		& 		& 4 & 0.362 & 0.087 & 0.087 & 0.237 & 0.054 & 0.055 & 0.229 & 0.051 & 0.052 \\ 
		& 		& 5 & 0.373 & 0.089 & 0.089 & 0.289 & 0.066 & 0.067 & 0.230 & 0.050 & 0.051 \\ 
		& 		& 6 & 0.458 & 0.111 & 0.112 & 0.364 & 0.086 & 0.087 & 0.270 & 0.069 & 0.069 \\ 
		& 		& 7 & 0.615 & 0.150 & 0.151 & 0.444 & 0.106 & 0.107 & 0.316 & 0.074 & 0.074 \\ 
		& 		& 8 & 0.645 & 0.158 & 0.158 & 0.438 & 0.104 & 0.105 & 0.281 & 0.065 & 0.065 \\ 
		& 		& 9 & 0.655 & 0.160 & 0.161 & 0.537 & 0.128 & 0.128 & 0.365 & 0.084 & 0.085 \\ 
		& 		& 10 & 0.300 & 0.072 & 0.072 & 1.001 & 0.243 & 0.243 & 0.215 & 0.086 & 0.088 \\ 
\cmidrule{3-12}
		& 		&  Mean & 0.432 & 0.104 & 0.105 & 0.394 & 0.093 & 0.093 & \textBF{0.259} & \textBF{0.063} & \textBF{0.064} \\ 
\midrule
Male & EVR & 1 & 1.246 & 0.308 & 0.309 & 0.643 & 0.157 & 0.158 & 0.242 & 0.057 & 0.058 \\ 
  	&	& 2 & 1.392 & 0.344 & 0.345 & 0.687 & 0.168 & 0.169 & 0.228 & 0.054 & 0.054 \\ 
	&	& 3 & 1.510 & 0.374 & 0.375 & 0.701 & 0.172 & 0.172 & 0.214 & 0.050 & 0.051 \\ 
  	&	& 4 & 1.632 & 0.403 & 0.405 & 0.757 & 0.185 & 0.186 & 0.301 & 0.071 & 0.072 \\ 
 	&	& 5 & 1.719 & 0.425 & 0.427 & 0.753 & 0.184 & 0.185 & 0.265 & 0.063 & 0.063 \\ 
	& 	& 6 & 1.908 & 0.472 & 0.474 & 0.901 & 0.221 & 0.221 & 0.410 & 0.098 & 0.098 \\ 
 	&	& 7 & 2.112 & 0.523 & 0.525 & 0.994 & 0.244 & 0.245 & 0.470 & 0.114 & 0.114 \\ 
 	&	& 8 & 2.115 & 0.524 & 0.526 & 0.905 & 0.222 & 0.222 & 0.434 & 0.105 & 0.105 \\ 
 	&	& 9 & 2.077 & 0.514 & 0.516 & 0.849 & 0.207 & 0.208 & 0.454 & 0.109 & 0.109 \\ 
 	&	& 10 & 1.250 & 0.309 & 0.309 & 0.552 & 0.131 & 0.132 & 0.187 & 0.044 & 0.044 \\ 
\cmidrule{3-12}
	&	& Mean & 1.696 & 0.419 & 0.421 & 0.774 & 0.189 & 0.190 & \textBF{0.321} & \textBF{0.076} & \textBF{0.077} \\
\cmidrule{2-12}
		& $K=6$ 	& 1 & 0.231 & 0.055 & 0.055 & 0.251 & 0.059 & 0.059 & 0.233 & 0.054 & 0.053 \\ 
	 	& 		&  2 & 0.241 & 0.057 & 0.058 & 0.249 & 0.058 & 0.058 & 0.220 & 0.051 & 0.051 \\ 
	 	& 		&  3 & 0.203 & 0.048 & 0.048 & 0.266 & 0.062 & 0.063 & 0.260 & 0.059 & 0.059 \\ 
	 	& 		&  4 & 0.262 & 0.062 & 0.063 & 0.312 & 0.073 & 0.073 & 0.331 & 0.077 & 0.078 \\ 
	 	& 		&  5 & 0.269 & 0.064 & 0.065 & 0.271 & 0.064 & 0.064 & 0.322 & 0.077 & 0.074 \\ 
	 	& 		&  6 & 0.355 & 0.085 & 0.086 & 0.423 & 0.100 & 0.101 & 0.480 & 0.114 & 0.114 \\ 
	 	& 		&  7 & 0.420 & 0.102 & 0.102 & 0.370 & 0.088 & 0.088 & 0.506 & 0.121 & 0.122 \\ 
	 	& 		&  8 & 0.401 & 0.097 & 0.098 & 0.378 & 0.090 & 0.090 & 0.518 & 0.124 & 0.125 \\ 
	 	& 		&  9 & 0.438 & 0.106 & 0.106 & 0.484 & 0.115 & 0.115 & 0.653 & 0.155 & 0.156 \\ 
	 	& 		&  10 & 0.223 & 0.053 & 0.053 & 0.293 & 0.067 & 0.068 & 0.288 & 0.067 & 0.067 \\ 
\cmidrule{3-12}
	 	& 		&  Mean & \textBF{0.304} & \textBF{0.073} & \textBF{0.073} & 0.330 & 0.077 & 0.078 & 0.381 & 0.090 & 0.090 \\
\end{longtable}
\end{center}
\end{singlespace}

For forecasting the Swedish female life-table death counts, we observe
\begin{inparaenum}
\item[1)] The CoDa method with the fitting period from 1950 onwards produces more accurate point forecasts than the one with all available data in the fitting period.
\item[2)] From a forecast accuracy perspective, it is advantageous to use $K=6$, compared with the eigenvalue ratio criterion.
\item[3)] Based on the averaged error criterion, the weighted CoDa method generally produces more accurate forecasts than the standard CoDa method. When $h=1, 2$, we also show that the CoDa method with the fitting period from 1950 onwards can outperform the weighted CoDa method. 
\end{inparaenum}

For forecasting the Swedish male life-table death counts, we observe
\begin{inparaenum}
\item[1)] The CoDa method with a longer fitting period produces more accurate point forecasts than the one from 1950 onwards. 
\item[2)] The weighted CoDa method produces more accurate point forecasts for shorter horizons rather than at longer horizons. 
\item[3)] From a forecast accuracy perspective, it is advantageous to use $K=6$, compared with the eigenvalue ratio criterion. Hereafter, we report the results based on $K=6$.
\end{inparaenum}

\section{Comparison of interval forecast accuracy}\label{sec:6}

In Table~\ref{tab:4_add}, we tabulate the estimated weight parameter in the weighted CoDa method under the CPD metrics. Notably, the selected weight parameters all lie between 0.051 and 0.147. Using the selected weight parameters in Table~\ref{tab:4_add}, we present interval forecast accuracy between the standard and weighted CoDa methods for the Swedish data. 
\begin{table}[!htb]
\centering
\tabcolsep 0.65in
\caption{\small Estimated geometrically decaying weight parameters in the weighted CoDa method under the CPD.}\label{tab:4_add}
\begin{tabular}{@{}lrrrr@{}}
\toprule
& \multicolumn{2}{c}{Female} & \multicolumn{2}{c}{Male} \\
$h$ 	& $\alpha=0.2$ & 0.05 & 0.2 & 0.05 \\ \midrule
1 	& 0.075 & 0.061 & 0.083 & 0.106 \\ 
2 	& 0.083 & 0.090 & 0.100 & 0.136 \\ 
3 	& 0.085 & 0.089 & 0.100 & 0.117 \\ 
4 	& 0.081 & 0.093 & 0.108 & 0.100 \\ 
5 	& 0.084 & 0.087 & 0.108 & 0.137 \\ 
6 	& 0.078 & 0.100 & 0.126 & 0.100 \\ 
7 	& 0.083 & 0.086 & 0.114 & 0.142 \\ 
8 	& 0.086 & 0.091 & 0.100 & 0.102 \\ 
9 	& 0.076 & 0.051 & 0.100 & 0.100 \\ 
10 	& 0.100 & 0.100 & 0.147 & 0.100 \\ 
\bottomrule
\end{tabular}
\end{table}

As shown in Table~\ref{tab:4}, the weighted CoDa method produces smaller CPD values than those obtained from the standard CoDa methods at $\alpha = 0.05$. For $\alpha=0.2$, the CoDa method with the fitting period from 1950 performs the best.
\begin{singlespace}
\begin{center}
\tabcolsep 0.19in
\begin{longtable}{@{}llrrrrrr@{}}
\caption{\small Comparison of the CPD between the CoDa and its weighted variant for forecasting the age-specific Swedish data.} \label{tab:4} \\
\toprule
& &  \multicolumn{2}{c}{CoDa} & \multicolumn{2}{c}{CoDa (1950)} & \multicolumn{2}{c}{Weighted CoDa} \\
$K$ & $h$ & $\alpha=0.2$ & $\alpha=0.05$ & $\alpha=0.2$ & $\alpha=0.05$ & $\alpha=0.2$ & $\alpha=0.05$ \\
\midrule
 Female 	& 1 & 0.167 & 0.046 & 0.030 & 0.035 & 0.053 & 0.007 \\ 
		& 2 & 0.180 & 0.050 & 0.034 & 0.027 & 0.054 & 0.034 \\ 
		& 3 & 0.188 & 0.049 & 0.045 & 0.041 & 0.038 & 0.025 \\ 
 		& 4 & 0.187 & 0.050 & 0.012 & 0.048 & 0.023 & 0.032 \\ 
 		& 5 & 0.183 & 0.050 & 0.054 & 0.037 & 0.030 & 0.015 \\ 
	 	& 6 & 0.195 & 0.050 & 0.058 & 0.040 & 0.022 & 0.027 \\ 
 		& 7 & 0.195 & 0.050 & 0.034 & 0.015 & 0.012 & 0.014 \\ 
 		& 8 & 0.194 & 0.050 & 0.070 & 0.013 & 0.029 & 0.008 \\ 
 		& 9 & 0.186 & 0.050 & 0.088 & 0.004 & 0.007 & 0.023 \\ 
 		& 10 & 0.191 & 0.050 & 0.011 & 0.004 & 0.178 & 0.049 \\ 
\cmidrule{2-8}
		& Mean  & 0.187 & 0.050 & 0.044 & 0.026 & 0.045 & 0.023 \\ 
\midrule
Male & 1 & 0.191 & 0.050 & 0.005 & 0.028 & 0.083 & 0.027 \\ 
	&  2 & 0.188 & 0.049 & 0.016 & 0.046 & 0.080 & 0.033 \\ 
 	&  3 & 0.195 & 0.050 & 0.045 & 0.022 & 0.057 & 0.005 \\ 
 	& 4 & 0.195 & 0.049 & 0.072 & 0.041 & 0.059 & 0.004 \\ 
 	& 5 & 0.195 & 0.050 & 0.009 & 0.025 & 0.036 & 0.024 \\ 
 	& 6 & 0.196 & 0.050 & 0.085 & 0.049 & 0.086 & 0.005 \\ 
 	& 7 & 0.195 & 0.050 & 0.041 & 0.047 & 0.036 & 0.011 \\ 
 	& 8 & 0.194 & 0.050 & 0.028 & 0.013 & 0.044 & 0.001 \\ 
 	& 9 & 0.200 & 0.050 & 0.057 & 0.049 & 0.025 & 0.000 \\ 
 	& 10 & 0.191 & 0.050 & 0.074 & 0.023 & 0.106 & 0.014 \\ 
\cmidrule{2-8}
 	& Mean & 0.194 & 0.050 & 0.043 & 0.034 & 0.061 & 0.012 \\ 
\bottomrule
\end{longtable}
\end{center}
\end{singlespace}

\section{\mbox{Application to a single-premium temporary immediate annuity}}\label{sec:7}

An important use of mortality forecasts for individuals over 60 is in the pension and insurance industries, whose profitability and solvency depend on accurate mortality forecasts to hedge longevity risk. Longevity risk is the chance that life expectancies exceed expectations for pricing, resulting in greater than anticipated cash flow needs from insurance companies or pension funds. When a person retires, an optimal way of guaranteeing one individual's financial income in retirement is to purchase an annuity \citep[see][]{Yaari65}. An annuity is a financial contract offered by insurers that guarantees a steady stream of income for a temporary or lifetime of the annuitants in exchange for an initial premium charge.

Following \cite{SH17}, we consider temporary annuities, which have grown in popularity in many developed countries, because immediate lifetime annuities, where rates are locked in for life, have been shown to deliver poor value for money \citep[see][Chapter 6]{CT08}. These temporary annuities pay a pre-determined and guaranteed income level higher than the income provided by a lifetime annuity for a similar premium. Fixed-term annuities offer a more affordable alternative to lifetime annuities and allow the purchaser to purchase a deferred annuity to address the tail longevity risk.

\begin{figure}[!htb]
\centering
\subfloat[CoDa]
{\includegraphics[width=8.4cm]{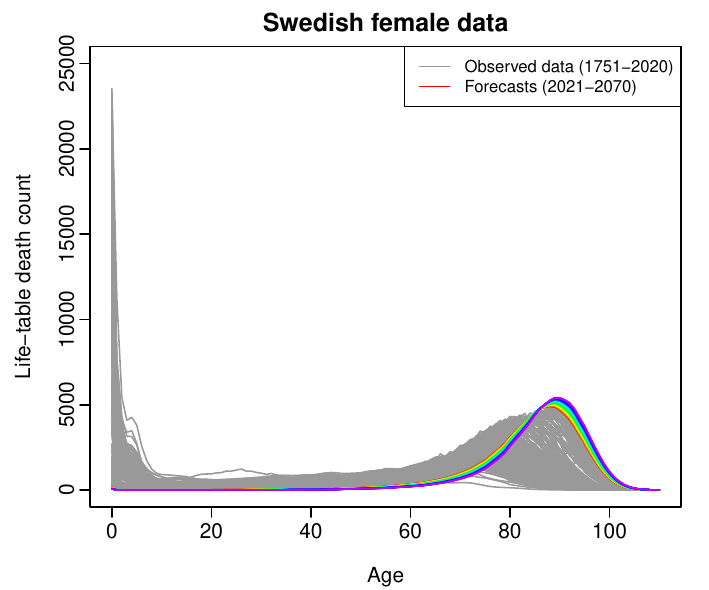}}
\quad
\subfloat[Weighted CoDa]
{\includegraphics[width=8.4cm]{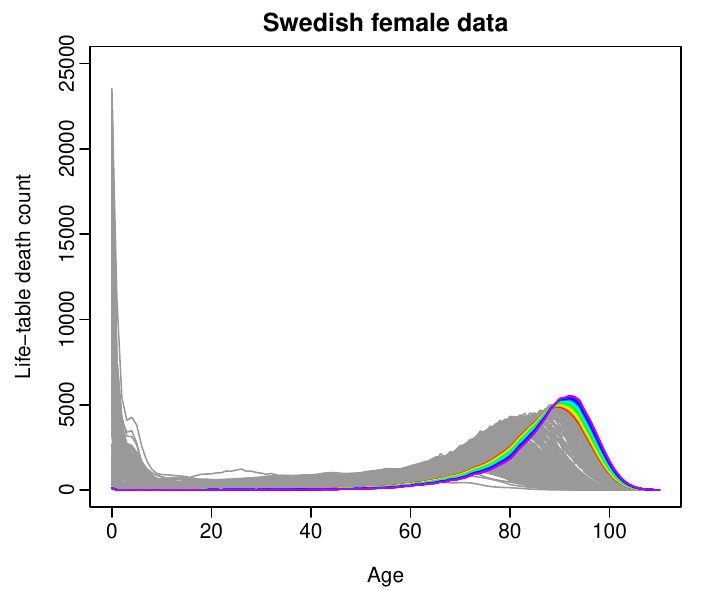}}
\\
\subfloat[CoDa]
{\includegraphics[width=8.4cm]{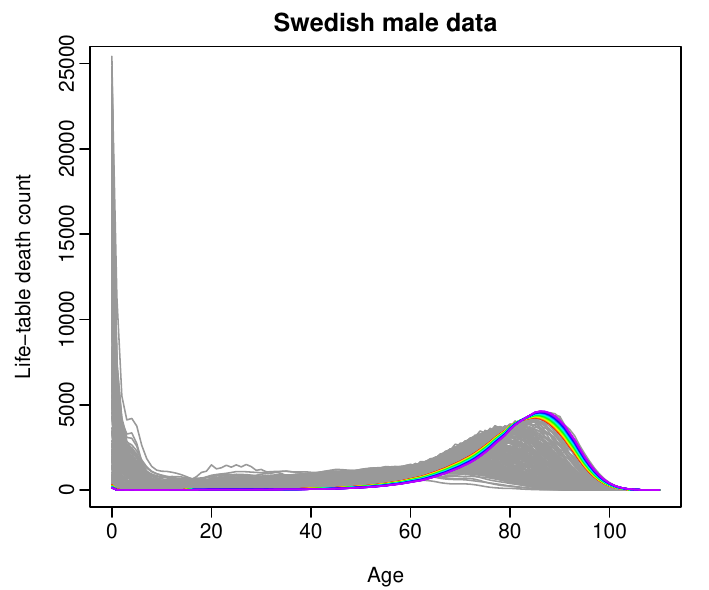}}
\quad
\subfloat[Weighted CoDa]
{\includegraphics[width=8.4cm]{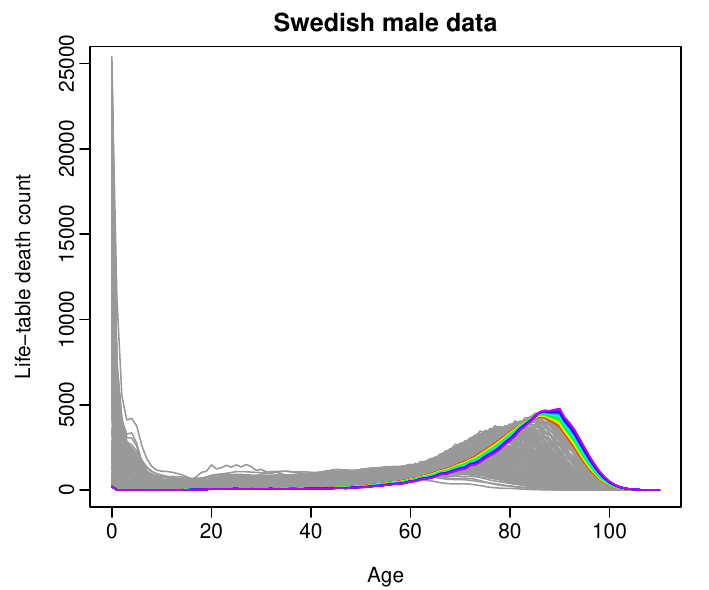}}
\caption{\small Age-specific life-table death count point forecasts from 2021 to 2070 for Swedish females and males.}\label{fig:5}
\end{figure}

We obtain forecasts of life-table death counts using the standard and weighted CoDa methods and then determine the corresponding survival probabilities. Using the RWD forecasting method, we display the forecasts of the life-table death counts from 2021 to 2070 for Swedish females and males in Figure~\ref{fig:5}.

With the forecasts of the life-table death counts, we compute a single-premium temporary immediate annuity \citep[see][p.114]{DHW09}, and we adopt a cohort approach to the computation of the survival probabilities. The $\tau$ year survival probability of a person aged $x$ at $t=0$ (or year $x$) is determined by
\[
_{\tau}p_x = \prod^{\tau-1}_{x=0}p_{x} = \prod^{\tau-1}_{x=1}(1-q_{x}) = \prod^{\tau-1}_{x=1}(1-\frac{d_{x}}{l_{x}}),
\]
where $d_{x}$ denotes the number of death counts between two successive ages $x$ and $x+1$, and $l_{x}$ denotes the number of lives alive at age $x$.

Annuity price with a maturity period of $T$ year is a random quantity, which depends on the value of zero-coupon bond price and future mortality. For an $x$-year-old with benefits \$1 per year, the temporary immediate annuity price can be written as
\[
a_x^T = \sum^T_{\tau=1}B(0,\tau)_{\tau}p_x,
\]
where $B(0,\tau)$ is the $\tau$-year bond price and $_{\tau}p_x$ denotes the survival probability.

In Table~\ref{tab:5}, we compute our best estimate of annuity prices for various ages and maturities for female and male policyholders residing in Sweden. We assume a constant interest rate at $\zeta=3\%$ and a zero-coupon bond is given as
\[
B(0,\tau) =\exp^{-\zeta\tau}.
\]

\vspace{-.4in}

\begin{singlespace}
\begin{center}
\tabcolsep 0.058in
\begin{longtable}{@{}lrrrrrrrrrrrr@{}}
\caption{\small Estimates of annuity prices with various ages and maturities $(T)$ for female and male policyholders residing in Sweden, when the interest rate is set at 3\%.}\label{tab:5}\\
\toprule
	& \multicolumn{6}{c}{CoDa} & \multicolumn{6}{c}{Weighted CoDa} \\
 Age & $T=5$ & 10 & 15 & 20 & 25 & 30 & 5 & 10 & 15 & 20 & 25 & 30 \\ 
\midrule
	\multicolumn{2}{l}{\hspace{-0.06in}{\underline{Female}}} & \\  
  60 & 4.517 & 8.289 & 11.385 & 13.844 & 15.662 & 16.791 & 4.515 & 8.276 & 11.354 & 13.785 & 15.571 & 16.684 \\ 
  65 & 4.484 & 8.165 & 11.089 & 13.250 & 14.592 & 15.177 & 4.476 & 8.139 & 11.032 & 13.157 & 14.482 & 15.068 \\ 
  70 & 4.433 & 7.953 & 10.555 & 12.171 & 12.876 & 13.045 & 4.423 & 7.917 & 10.484 & 12.084 & 12.792 & 12.967 \\ 
  75 & 4.334 & 7.537 & 9.526 & 10.394 & 10.603 & 10.624 & 4.319 & 7.490 & 9.469 & 10.343 & 10.559 & 10.582 \\ 
  80 & 4.116 & 6.673 & 7.788 & 8.057 & 8.084 & 8.085 & 4.103 & 6.662 & 7.793 & 8.072 & 8.102 & 8.103 \\ 
  85 & 3.646 & 5.236 & 5.619 & 5.657 & 5.659 &  & 3.665 & 5.285 & 5.686 & 5.728 & 5.730 &  \\ 
  90 & 2.894 & 3.590 & 3.660 & 3.662 &  &  & 2.921 & 3.643 & 3.720 & 3.723 &  &  \\ 
  95 & 2.036 & 2.241 & 2.248 &  &  &  & 2.079 & 2.300 & 2.309 &  &  &  \\ 
  100 & 1.323 & 1.370 &  &  &  &  & 1.359 & 1.412 &  &  &  &  \\ 
  105 & 0.875 &  &  &  &  &  & 0.907 &  &  &  &  &  \\ 
\\
	\multicolumn{2}{l}{\hspace{-0.06in}{\underline{Male}}} & \\  
  60 & 4.485 & 8.161 & 11.082 & 13.278 & 14.742 & 15.502 & 4.486 & 8.166 & 11.097 & 13.305 & 14.782 & 15.554 \\ 
  65 & 4.427 & 7.946 & 10.591 & 12.354 & 13.269 & 13.569 & 4.431 & 7.959 & 10.617 & 12.396 & 13.325 & 13.631 \\ 
  70 & 4.339 & 7.600 & 9.774 & 10.903 & 11.272 & 11.332 & 4.344 & 7.617 & 9.807 & 10.952 & 11.329 & 11.391 \\ 
  75 & 4.176 & 6.959 & 8.404 & 8.877 & 8.953 & 8.958 & 4.181 & 6.979 & 8.441 & 8.923 & 9.002 & 9.007 \\ 
  80 & 3.838 & 5.831 & 6.483 & 6.588 & 6.595 & 6.595 & 3.851 & 5.864 & 6.527 & 6.635 & 6.642 & 6.642 \\ 
  85 & 3.249 & 4.312 & 4.484 & 4.495 & 4.495 &  & 3.260 & 4.334 & 4.509 & 4.520 & 4.520 &  \\ 
  90 & 2.438 & 2.831 & 2.856 & 2.857 &  &  & 2.449 & 2.846 & 2.873 & 2.873 &  &  \\ 
  95 & 1.671 & 1.780 & 1.783 &  &  &  & 1.680 & 1.790 & 1.792 &  &  &  \\ 
  100 & 1.110 & 1.137 &  &  &  &  & 1.112 & 1.139 &  &  &  &  \\ 
  105 & 0.777 &  &  &  &  &  & 0.780 &  &  &  &  &  \\ 
\bottomrule
\end{longtable}
\end{center}
\end{singlespace}

For the male population, the annuity prices for the weighted CoDa method are higher than the ones produced from the CoDa method. This may indicate the increasing life expectancy for the male population. For the female population, the weighted CoDa method recognizes the longevity risk trend more than its unweighted counterpart; thus, it produces different mortality forecasts. Between ages 60 and 80, the weighted CoDa mortality forecasts are predicted to be lower than the unweighted counterpart. This implies fewer people are likely to die in that age group. For ages 85 and above, this is a longevity risk group with increasingly more people.

To provide forecast uncertainty, we rely on the bootstrapped life-table death counts, derive the survival probabilities, and compute the associated annuities for different ages and maturities. Since we consider ages from 60 to 110, we construct 50 steps ahead bootstrap forecasts of life-table death counts. In Table~\ref{tab:7}, we present the 95\% pointwise prediction intervals of annuities for different ages and maturities, where age + maturity $\leq 110$.	 

\vspace{-.3in}	 
\begin{singlespace}
\begin{small}
\begin{center}
\tabcolsep 0.05in
\begin{longtable}{@{}llcccccc@{}}
\caption{\small Ninety-five percentage pointwise prediction intervals of annuity prices with different ages and maturities ($T$) for female and male policyholders residing in Sweden, when the interest rate is set at 3\%.}\label{tab:7}\\
\toprule
 Sex & Age & $T=5$ & $T=10$ & $T=15$ & $T=20$ & $T=25$ & $T=30$ \\ 
\midrule
	\multicolumn{2}{l}{\hspace{-0.06in}{\underline{CoDa}}} & \\
F & 60 & (4.487, 4.533) & (8.206, 8.338) & (11.235, 11.492) & (13.652, 14.028) & (15.438, 15.951) & (16.556, 17.220) \\ 
   & 65 & (4.433, 4.514) & (8.056, 8.254) & (10.913, 11.267) & (13.037, 13.554) & (14.344, 15.073) & (14.924, 15.804) \\ 
   & 70 & (4.355, 4.487) & (7.793, 8.114) & (10.329, 10.874) & (11.931, 12.689) & (12.593, 13.584) & (12.769, 13.824) \\ 
   & 75 & (4.247, 4.426) & (7.360, 7.817) & (9.333, 10.059) & (10.174, 11.154) & (10.382, 11.435) & (10.409, 11.476) \\ 
   & 80 & (3.991, 4.287) & (6.460, 7.167) & (7.516, 8.569) & (7.751, 8.947) & (7.788, 9.000) & (7.789, 9.002) \\ 
   & 85 & (3.476, 3.949) & (4.932, 5.940) & (5.284, 6.500) & (5.321, 6.580) & (5.321, 6.583) &   \\ 
   & 90 & (2.620, 3.399) & (3.203, 4.376) & (3.267, 4.477) & (3.270, 4.490) &  &   \\ 
   & 95 & (1.707, 2.575) & (1.878, 2.923) & (1.884, 2.935) &  &  &  \\ 
   & 100 & (0.985, 1.876) & (1.014, 1.974) &  &  &  &   \\ 
   & 105 & (0.597, 1.339) &  &  &  &  &    \\ 
\\
M & 60 & (4.435, 4.510) & (8.032, 8.239) & (10.875, 11.235) & (12.972, 13.522) & (14.356, 15.125) & (15.083, 16.013) \\ 
    & 65 & (4.359, 4.477) & (7.792, 8.072) & (10.314, 10.840) & (11.989, 12.758) & (12.866, 13.841) & (13.185, 14.250) \\ 
    & 70 & (4.231, 4.419) & (7.330, 7.822) & (9.397, 10.193) & (10.471, 11.529) & (10.841, 12.021) & (10.907, 12.108) \\ 
    & 75 & (4.029, 4.307) & (6.689, 7.336) & (8.064, 9.028) & (8.527, 9.679) & (8.615, 9.792) & (8.619, 9.799) \\ 
    & 80 & (3.653, 4.073) & (5.524, 6.387) & (6.143, 7.247) & (6.251, 7.420) & (6.256, 7.428) & (6.256, 7.428) \\ 
    & 85 & (3.050, 3.626) & (4.017, 4.996) & (4.182, 5.267) & (4.193, 5.295) & (4.193, 5.296) &    \\ 
    & 90 & (2.121, 2.916) & (2.460, 3.512) & (2.484, 3.556) & (2.485, 3.558) &  &  \\ 
    & 95 & (1.383, 2.171) & (1.467, 2.361) & (1.468, 2.367) &  &  &  \\ 
    & 100 & (0.769, 1.611) & (0.788, 1.665) &  &  &  &   \\ 
    & 105 & (0.508, 1.220) &  &  &  &  &   \\ 

\\
\multicolumn{4}{l}{\hspace{-0.06in}{\underline{Weighted CoDa}}} & \\
F & 60 & (4.439, 4.521) & (8.048, 8.268) & (10.904, 11.318) & (13.100, 13.715) & (14.752, 15.548) & (15.881, 16.836) \\ 
   & 65 & (4.366, 4.488) & (7.813, 8.130) & (10.476, 11.016) & (12.469, 13.234) & (13.829, 14.795) & (14.517, 15.672) \\ 
   & 70 & (4.252, 4.444) & (7.498, 7.982) & (9.980, 10.708) & (11.632, 12.622) & (12.502, 13.719) & (12.751, 14.078) \\ 
   & 75 & (4.198, 4.404) & (7.323, 7.829) & (9.422, 10.264) & (10.507, 11.658) & (10.836, 12.116) & (10.877, 12.180) \\ 
   & 80 & (4.028, 4.353) & (6.725, 7.514) & (8.103, 9.312) & (8.512, 9.937) & (8.573, 10.024) & (8.575, 10.031) \\ 
   & 85 & (3.676, 4.189) & (5.541, 6.616) & (6.095, 7.418) & (6.166, 7.547) & (6.168, 7.552) &   \\ 
   & 90 & (2.891, 3.607) & (3.746, 4.848) & (3.858, 5.048) & (3.865, 5.057) &  &   \\ 
   & 95 & (2.094, 2.757) & (2.355, 3.205) & (2.368, 3.228) &  &  & \\ 
   & 100 & (1.321, 1.813) & (1.375, 1.907) &  &  &  &  \\ 
   & 105 & (0.821, 1.148) &  &  &  &  &    \\ 
\\
M & 60 & (4.423, 4.515) & (8.032, 8.267) & (10.936, 11.351) & (13.218, 13.833) & (14.859, 15.712) & (15.863, 16.939) \\ 
    & 65 & (4.371, 4.496) & (7.883, 8.202) & (10.612,11.177) & (12.611, 13.443) & (13.781, 14.930) & (14.273, 15.607) \\ 
    & 70 & (4.300, 4.488) & (7.651, 8.128) & (10.030, 10.908) & (11.500, 12.751) & (12.061, 13.543) & (12.187, 13.713) \\ 
    & 75 & (4.146, 4.443) & (7.102, 7.877) & (8.908, 10.119) & (9.649, 11.116) & (9.767, 11.342) & (9.777, 11.361) \\ 
    & 80 & (3.841, 4.330) & (6.119, 7.239) & (6.996, 8.562) & (7.192, 8.846) & (7.207, 8.872) & (7.207, 8.873) \\ 
    & 85 & (3.231, 3.936) & (4.483, 5.749) & (4.736, 6.159) & (4.756, 6.191) & (4.757, 6.192) &  \\ 
    & 90 & (2.411, 3.123) & (2.891, 3.846) & (2.926, 3.900) & (2.927, 3.900) &  &  \\ 
    & 95 & (1.640, 2.125) & (1.747, 2.291) & (1.749, 2.296) &  &  &  \\ 
    & 100 & (0.953, 1.303) & (0.975, 1.336) &  &  &  &   \\ 
    & 105 & (0.582, 0.896) &  &  &  &  &    \\ 
\bottomrule
\end{longtable}
\end{center}
\end{small}
\end{singlespace}

\vspace{-.2in}

Appendix~\ref{sec:app_A} presents the point estimate of annuity prices for various ages and maturities when the interest rate $\zeta=1\%$ and 5\%, respectively. In Appendix~\ref{sec:app_B}, we show the 95\% pointwise prediction intervals.

\section{Conclusion}\label{sec:8}

We present an extension of the CoDa method by incorporating geometrically decaying weights to estimate mean function and functional principal components. The weighted CoDa method can improve the forecast accuracy of age-specific life-table death counts, which could help actuaries enhance the accuracy of their pricing of annuities and setting of reserves.

In the weighted CoDa method, larger weights are assigned to the most recent data, whereas the data from the distant past are less important to forecasting. We select the estimated optimal value of the weight parameter by minimizing the KLD and two variants of the JSD using a set of validation data. We compare the point forecast errors between the standard and weighted CoDa methods in the testing data with the estimated weight parameter. The weighted CoDa method generally improves accuracy compared with its unweighted counterparts. 

From the viewpoint of forecast accuracy, we suggest implementing the weighted CoDa method. From the perspective of actuarial calculations, the improvement in mortality leads to a more accurate estimate of annuity prices. To facilitate reproducibility, the \Rlogo \ code for implementing all the methods is available at \url{https://github.com/hanshang/Weighted_CoDa}.

There are a few ways in which the present paper can be further extended, and we briefly mention five: 
\begin{inparaenum}
\item[1)] Although we demonstrate the practicability of the proposal via the Swedish data, the weighted CoDa method can be applied to other countries (such as Denmark, Japan, and the USA), especially with long-run high-quality mortality data. 
\item[2)] One could implement a hypothesis test, such as the Friedman and Nemenyi tests in \cite{Shang15}, to examine the statistical significance between the standard and weighted CoDa methods.
\item[3)] One could study the uncertainty associated with the estimated weight parameter.
\item[4)] One could apply the weighted CoDa method to death counts from cohort-based life tables.
\item[5)] Centered log-ratio transformation is one of many possible transformations. We may also study additive log-ratio transformation in \cite{Aitchison86}, square-root transformation in \cite{SW11} or $\alpha$ transformation in \cite{TS20}.
\end{inparaenum}

\section*{Acknowledgment}

The authors thank a reviewer for insightful comments and suggestions. The first author acknowledges financial support from an Australian Research Council Discovery Project DP230102250.

\newpage
\begin{appendices} 
\numberwithin{equation}{section}
\section{Estimate of annuity prices when interest rate $\zeta=1\%$ or 5\%}\label{sec:app_A}

In Tables~\ref{tab:8}, we present the estimated annuity premium prices with \$1 benefit for various ages and maturities when the interest rate $\zeta=1\%$.
\begin{table}[!htb]
\centering
\tabcolsep 0.062in
\caption{\small Estimates of annuity prices with various ages and maturities $(T)$ for female and male policyholders residing in Sweden, when the interest rate is set at 1\%.}\label{tab:8}
\begin{tabular}{@{}lrrrrrrrrrrrr@{}}
\toprule
& \multicolumn{6}{c}{CoDa} & \multicolumn{6}{c}{Weighted CoDa} \\
 Age & $T=5$ & 10 & 15 & 20 & 25 & 30 & 5 & 10 & 15 & 20 & 25 & 30 \\ 
\midrule
\multicolumn{2}{l}{\hspace{-0.06in}{\underline{Female}}} & \\
60 	& 4.791 & 9.213 & 13.224 & 16.743 & 19.615 & 21.582 & 4.789 & 9.198 & 13.184 & 16.663 & 19.485 & 21.426 \\ 
65 	& 4.756 & 9.071 & 12.856 & 15.945 & 18.061 & 19.078 & 4.748 & 9.040 & 12.786 & 15.824 & 17.914 & 18.931 \\ 
70 	& 4.701 & 8.825 & 12.191 & 14.496 & 15.604 & 15.896 & 4.690 & 8.784 & 12.104 & 14.388 & 15.499 & 15.802 \\ 
75 	& 4.594 & 8.344 & 10.912 & 12.146 & 12.472 & 12.508 & 4.578 & 8.291 & 10.845 & 12.088 & 12.425 & 12.465 \\ 
80 	& 4.360 & 7.346 & 8.781 & 9.160 & 9.202 & 9.204 & 4.345 & 7.335 & 8.790 & 9.185 & 9.231 & 9.233 \\ 
85 	& 3.853 & 5.705 & 6.194 & 6.248 & 6.250 &  & 3.874 & 5.760 & 6.272 & 6.332 & 6.335 &  \\ 
90 	& 3.047 & 3.853 & 3.942 & 3.945 &  &  & 3.076 & 3.912 & 4.010 & 4.014 &  &  \\ 
95 	& 2.132 & 2.367 & 2.377 &  &  &  & 2.178 & 2.432 & 2.443 &  &  &  \\ 
100 	& 1.377 & 1.430 &  &  &  &  & 1.415 & 1.475 &  &  &  &  \\ 
105 	& 0.906 &  &  &  &  &  & 0.940 &  &  &  &  &  \\ 
\\
\multicolumn{2}{l}{\hspace{-0.06in}{\underline{Male}}} & \\    
60 	& 4.757 & 9.065 & 12.848 & 15.989 & 18.299 & 19.620 & 4.758 & 9.072 & 12.866 & 16.023 & 18.355 & 19.698 \\ 
65 	& 4.695 & 8.818 & 12.241 & 14.758 & 16.199 & 16.718 & 4.699 & 8.833 & 12.272 & 14.812 & 16.275 & 16.806 \\ 
70 	& 4.600 & 8.419 & 11.227 & 12.835 & 13.414 & 13.516 & 4.606 & 8.438 & 11.268 & 12.898 & 13.489 & 13.595 \\ 
75 	& 4.424 & 7.678 & 9.540 & 10.211 & 10.329 & 10.338 & 4.430 & 7.702 & 9.586 & 10.269 & 10.391 & 10.400 \\ 
80 	& 4.061 & 6.384 & 7.221 & 7.369 & 7.380 & 7.380 & 4.074 & 6.421 & 7.272 & 7.424 & 7.435 & 7.436 \\ 
85 	& 3.427 & 4.662 & 4.880 & 4.896 & 4.896 &  & 3.439 & 4.686 & 4.909 & 4.925 & 4.925 &  \\ 
90 	& 2.560 & 3.013 & 3.046 & 3.046 &  &  & 2.571 & 3.031 & 3.064 & 3.065 &  &  \\ 
95 	& 1.745 & 1.870 & 1.873 &  &  &  & 1.754 & 1.880 & 1.883 &  &  &  \\ 
100 	& 1.152 & 1.183 &  &  &  &  & 1.155 & 1.186 &  &  &  &  \\ 
105 	& 0.804 &  &  &  &  &  & 0.807 &  &  &  &  &  \\ 
\bottomrule
\end{tabular}
\end{table}

\newpage

In Tables~\ref{tab:9}, we present the estimated annuity premium prices with \$1 benefit for various ages and maturities when the interest rate $\zeta=5\%$.

\begin{table}[!htb]
\centering
\tabcolsep 0.068in
\caption{\small Estimates of annuity prices with various ages and maturities $(T)$ for female and male policyholders residing in Sweden, when the interest rate is set at 5\%.}\label{tab:9}
\begin{tabular}{@{}lrrrrrrrrrrrr@{}}
\toprule
& \multicolumn{6}{c}{CoDa} & \multicolumn{6}{c}{Weighted CoDa} \\
Age & $T=5$ & 10 & 15 & 20 & 25 & 30 & 5 & 10 & 15 & 20 & 25 & 30 \\ 
\midrule
\multicolumn{2}{l}{\hspace{-0.06in}{\underline{Female}}} & \\   
60 	& 4.261 & 7.482 & 9.874 & 11.594 & 12.745 & 13.394 & 4.259 & 7.471 & 9.849 & 11.549 & 12.681 & 13.320 \\ 
65 	& 4.231 & 7.375 & 9.634 & 11.147 & 11.998 & 12.336 & 4.224 & 7.351 & 9.588 & 11.075 & 11.916 & 12.254 \\ 
70 	& 4.183 & 7.190 & 9.203 & 10.336 & 10.786 & 10.884 & 4.174 & 7.159 & 9.145 & 10.268 & 10.718 & 10.819 \\ 
75 	& 4.091 & 6.830 & 8.372 & 8.983 & 9.117 & 9.129 & 4.077 & 6.789 & 8.323 & 8.938 & 9.076 & 9.090 \\ 
80 	& 3.889 & 6.080 & 6.947 & 7.138 & 7.155 & 7.155 & 3.877 & 6.069 & 6.949 & 7.147 & 7.166 & 7.167 \\ 
85 	& 3.452 & 4.819 & 5.119 & 5.146 & 5.147 &  & 3.470 & 4.863 & 5.176 & 5.206 & 5.207 &  \\ 
90 	& 2.750 & 3.352 & 3.407 & 3.409 &  &  & 2.775 & 3.399 & 3.460 & 3.462 &  &  \\ 
95 	& 1.946 & 2.124 & 2.130 &  &  &  & 1.986 & 2.179 & 2.186 &  &  &  \\ 
100 	& 1.272 & 1.313 &  &  &  &  & 1.306 & 1.353 &  &  &  &  \\ 
105 	& 0.846 &  &  &  &  &  & 0.876 &  &  &  &  &  \\ 
\\
\multicolumn{2}{l}{\hspace{-0.06in}{\underline{Male}}} & \\      
60 	& 4.232 & 7.370 & 9.629 & 11.165 & 12.094 & 12.531 & 4.233 & 7.376 & 9.640 & 11.185 & 12.122 & 12.566 \\ 
65 	& 4.178 & 7.184 & 9.230 & 10.465 & 11.047 & 11.220 & 4.182 & 7.195 & 9.251 & 10.497 & 11.088 & 11.265 \\ 
70 	& 4.096 & 6.884 & 8.567 & 9.360 & 9.596 & 9.631 & 4.101 & 6.899 & 8.595 & 9.399 & 9.640 & 9.676 \\ 
75 	& 3.944 & 6.327 & 7.449 & 7.783 & 7.832 & 7.835 & 3.950 & 6.344 & 7.480 & 7.820 & 7.870 & 7.873 \\ 
80 	& 3.631 & 5.341 & 5.850 & 5.925 & 5.929 & 5.929 & 3.643 & 5.370 & 5.888 & 5.964 & 5.969 & 5.969 \\ 
85 	& 3.081 & 3.999 & 4.133 & 4.141 & 4.141 &  & 3.092 & 4.018 & 4.155 & 4.163 & 4.163 &  \\ 
90 	& 2.324 & 2.664 & 2.684 & 2.684 &  &  & 2.333 & 2.678 & 2.699 & 2.699 &  &  \\ 
95 	& 1.602 & 1.697 & 1.699 &  &  &  & 1.610 & 1.706 & 1.708 &  &  &  \\ 
100 	& 1.070 & 1.093 &  &  &  &  & 1.072 & 1.096 &  &  &  &  \\ 
105 	& 0.752 &  &  &  &  &  & 0.755 &  &  &  &  &  \\ 
\bottomrule
\end{tabular}
\end{table}

\newpage
\section{Ninety-five percentage prediction intervals of annuity prices when interest rate $\zeta=1\%$ or 5\%}\label{sec:app_B}

\begin{table}[!htb]
\centering
\tabcolsep 0.058in
\caption{\small Ninety-five percentage pointwise prediction intervals of annuity prices with different ages and maturities ($T$), when the interest rate $\zeta=1\%$.}\label{tab:10}
\begin{small}
\begin{tabular}{@{}llcccccc@{}}
\toprule
 Sex & Age & $T=5$ & $T=10$ & $T=15$ & $T=20$ & $T=25$ & $T=30$ \\ 
\midrule
\multicolumn{2}{l}{\hspace{-0.06in}{\underline{CoDa}}} & \\
F 	& 60 & (4.760, 4.809) & (9.117, 9.270) & (13.043, 13.355) & (16.497, 16.990) & (19.302, 20.034) & (21.240, 22.255) \\ 
	&  65 & (4.702, 4.789) & (8.946, 9.172) & (12.645, 13.075) & (15.662, 16.349) & (17.739, 18.760) & (18.735, 20.052) \\ 
	&  70 & (4.619, 4.760) & (8.647, 9.011) & (11.923, 12.584) & (14.180, 15.189) & (15.251, 16.593) & (15.550, 17.013) \\ 
	&  75 & (4.502, 4.694) & (8.144, 8.665) & (10.668, 11.564) & (11.871, 13.124) & (12.190, 13.577) & (12.248, 13.648) \\ 
	&  80 & (4.226, 4.543) & (7.102, 7.911) & (8.460, 9.721) & (8.800, 10.257) & (8.842, 10.349) & (8.844, 10.352) \\ 
	&  85 & (3.670, 4.179) & (5.371, 6.504) & (5.813, 7.224) & (5.857, 7.335) & (5.859, 7.341) &  \\ 
	&  90 & (2.759, 3.582) & (3.424, 4.719) & (3.508, 4.866) & (3.510, 4.880) &  &  \\ 
	&  95 & (1.779, 2.702) & (1.989, 3.106) & (1.995, 3.135) &  &  &   \\ 
	&  100 & (1.020, 1.960) & (1.054, 2.077) &  &  &  &  \\ 
	&  105 & (0.615, 1.390) &  &  &  &  & \\ 
  \\
M 	& 60 & (4.703, 4.784) & (8.917, 9.156) & (12.600, 13.036) & (15.592,16.314) & (17.762, 18.842) & (19.019, 20.410) \\ 
  	& 65 & (4.622, 4.748) & (8.642, 8.964) & (11.901, 12.549) & (14.288, 15.295) & (15.662, 17.007) & (16.202, 17.703) \\ 
  	& 70 & (4.484, 4.685) & (8.108, 8.673) & (10.775, 11.744) & (12.296, 13.657) & (12.901, 14.435) & (13.006, 14.583) \\ 
  	& 75 & (4.268, 4.565) & (7.374, 8.110) & (9.141, 10.297) & (9.786, 11.221) & (9.934, 11.404) & (9.942, 11.419) \\ 
  	& 80 & (3.864, 4.310) & (6.043, 7.010) & (6.851, 8.127) & (6.988, 8.382) & (7.005, 8.392) & (7.006, 8.393) \\ 
  	& 85 & (3.212, 3.831) & (4.333, 5.434) & (4.557, 5.778) & (4.574, 5.810) & (4.575, 5.811) &  \\ 
  	& 90 & (2.226, 3.070) & (2.621, 3.766) & (2.651, 3.821) & (2.652, 3.824) &  &  \\ 
  	& 95 & (1.443, 2.271) & (1.531, 2.493) & (1.533, 2.500) &  &  &   \\ 
  	& 100 & (0.800, 1.675) & (0.818, 1.748) &  &  &  &   \\ 
  	& 105 & (0.526, 1.266) &  &  &  &  &    \\ 
\\
\multicolumn{4}{l}{\hspace{-0.06in}{\underline{Weighted CoDa}}} & \\
F 	& 60 & (4.707, 4.796) & (8.933, 9.188) & (12.634, 13.139) & (15.768, 16.565) & (18.379, 19.471) & (20.358, 21.736) \\ 
  	& 65 & (4.629, 4.760) & (8.666, 9.028) & (12.108, 12.771) & (14.958, 15.943) & (17.102, 18.414) & (18.279, 19.953) \\ 
  	& 70 & (4.505, 4.712) & (8.309, 8.859) & (11.509, 12.384) & (13.868, 15.142) & (15.247, 16.873) & (15.680, 17.491) \\ 
  	& 75 & (4.448, 4.670) & (8.108, 8.682) & (10.810, 11.841) & (12.348, 13.840) & (12.812, 14.548) & (12.878, 14.662) \\ 
  	& 80 & (4.264, 4.614) & (7.414, 8.319) & (9.189, 10.639) & (9.763, 11.526) & (9.857, 11.672) & (9.861, 11.681) \\ 
  	& 85 & (3.888, 4.438) & (6.055, 7.275) & (6.759, 8.301) & (6.855, 8.469) & (6.860, 8.478) &   \\ 
  	& 90 & (3.048, 3.807) & (4.032, 5.254) & (4.185, 5.506) & (4.195, 5.521) &  & \\ 
  	& 95 & (2.195, 2.895) & (2.496, 3.417) & (2.510, 3.443) &  &  & \\ 
  	& 100 & (1.375, 1.888) & (1.439, 1.995) &  &  &  &   \\ 
  	& 105 & (0.849, 1.190) &  &  &  &  &    \\   
\\
M 	& 60 & (4.691, 4.789) & (8.921, 9.187) & (12.681, 13.186) & (15.937, 16.740) & (18.526, 19.720) & (20.259, 21.869) \\ 
  	& 65 & (4.635, 4.769) & (8.747, 9.113) & (12.281, 12.964) & (15.121, 16.225) & (16.966, 18.577) & (17.788, 19.747) \\ 
  	& 70 & (4.560, 4.760) & (8.478, 9.028) & (11.568, 12.631) & (13.621, 15.263) & (14.541, 16.518) & (14.746, 16.825) \\ 
  	& 75 & (4.396, 4.712) & (7.856, 8.733) & (10.180, 11.643) & (11.176, 13.079) & (11.373, 13.452) & (11.388, 13.466) \\ 
  	& 80 & (4.063, 4.589) & (6.715, 7.992) & (7.861, 9.695) & (8.124, 10.106) & (8.145, 10.135) & (8.145, 10.136) \\ 
  	& 85 & (3.410, 4.165) & (4.870, 6.267) & (5.188, 6.802) & (5.217, 6.849) & (5.218, 6.850) &   \\ 
  	& 90 & (2.533, 3.287) & (3.089, 4.128) & (3.133, 4.193) & (3.134, 4.195) &  &  \\ 
  	& 95 & (1.713, 2.226) & (1.835, 2.414) & (1.838, 2.419) &  &  &  \\ 
  	& 100 & (0.988, 1.354) & (1.012, 1.391) &  &  &  &    \\ 
  	& 105 & (0.601, 0.926) &  &  &  &  &   \\ 
\bottomrule
\end{tabular}
\end{small}
\end{table}

In Tables~\ref{tab:10} and~\ref{tab:11}, we present the pointwise prediction intervals of annuity prices with different ages and maturities ($T$) for female and male policyholders residing in Sweden when the interest rate $\zeta=1\%$ and $5\%$, respectively.
\begin{table}[!htb]
\centering
\tabcolsep 0.062in
\caption{\small Ninety-five percentage pointwise prediction intervals of annuity prices with different ages and maturities ($T$) when the interest rate $\zeta=5\%$.}\label{tab:11}
\begin{small}
\begin{tabular}{@{}llcccccc@{}}
\toprule
 Sex & Age & $T=5$ & $T=10$ & $T=15$ & $T=20$ & $T=25$ & $T=30$ \\ 
\midrule
	\multicolumn{2}{l}{\hspace{-0.06in}{\underline{CoDa}}} & \\
 F 	& 60 & (4.234, 4.277) & (7.410, 7.525) & (9.749, 9.961) & (11.443, 11.733) & (12.574, 12.946) & (13.222, 13.677) \\ 
 	&  65 & (4.183, 4.259) & (7.278, 7.452) & (9.488, 9.779) & (10.979, 11.378) & (11.808, 12.343) & (12.145, 12.757) \\ 
 	&  70 & (4.109, 4.234) & (7.048, 7.331) & (9.020, 9.466) & (10.130, 10.732) & (10.570, 11.303) & (10.678, 11.442) \\ 
 	&  75 & (4.010, 4.178) & (6.672, 7.073) & (8.208, 8.805) & (8.797, 9.581) & (8.936, 9.757) & (8.951, 9.778) \\ 
 	&  80 & (3.773, 4.050) & (5.890, 6.512) & (6.713, 7.603) & (6.879, 7.874) & (6.900, 7.899) & (6.900, 7.899) \\ 
 	&  85 & (3.294, 3.735) & (4.542, 5.438) & (4.818, 5.882) & (4.845, 5.928) & (4.845, 5.930) &   \\ 
 	&  90 & (2.490, 3.228) & (3.004, 4.066) & (3.054, 4.148) & (3.056, 4.154) &  &   \\ 
 	&  95 & (1.636, 2.456) & (1.783, 2.757) & (1.788, 2.765) &  &  &   \\ 
 	&  100 & (0.950, 1.796) & (0.976, 1.880) &  &  &  &  \\ 
 	&  105 & (0.579, 1.292) &  &  &  &  &    \\ 
  \\
  M   & 60 & (4.185, 4.255) & (7.258, 7.438) & (9.454, 9.753) & (10.932, 11.352) & (11.819, 12.365) & (12.224, 12.870) \\ 
 	& 65 & (4.115, 4.224) & (7.047, 7.293) & (9.004, 9.435) & (10.180, 10.771) & (10.740, 11.457) & (10.920, 11.686) \\ 
 	& 70 & (3.995, 4.170) & (6.653, 7.078) & (8.246, 8.908) & (9.002, 9.844) & (9.245, 10.158) & (9.279, 10.215) \\ 
 	& 75 & (3.806, 4.067) & (6.086, 6.657) & (7.161, 7.970) & (7.490, 8.419) & (7.549, 8.494) & (7.553, 8.499) \\ 
 	& 80 & (3.457, 3.850) & (5.067, 5.837) & (5.554, 6.504) & (5.623, 6.616) & (5.628, 6.621) & (5.628, 6.621) \\ 
 	& 85 & (2.895, 3.436) & (3.728, 4.610) & (3.856, 4.819) & (3.863, 4.839) & (3.864, 4.840) &  \\ 
 	& 90 & (2.024, 2.774) & (2.318, 3.290) & (2.333, 3.321) & (2.333, 3.325) &  &  \\ 
 	& 95 & (1.325, 2.077) & (1.402, 2.242) & (1.406, 2.248) &  &  &   \\ 
 	& 100 & (0.740, 1.546) & (0.761, 1.598) &  &  &  &  \\ 
 	& 105 & (0.491, 1.177) &  &  &  &  &   \\ 
  \\
  	\multicolumn{4}{l}{\hspace{-0.06in}{\underline{Weighted CoDa}}} & \\
F 	& 60 & (4.189, 4.266) & (7.272, 7.465) & (9.481, 9.822) & (11.019, 11.495) & (12.067, 12.654) & (12.712, 13.389) \\ 
 	& 65 & (4.122, 4.234) & (7.068, 7.344) & (9.130, 9.571) & (10.526, 11.125) & (11.388, 12.108) & (11.801, 12.613) \\ 
 	& 70 & (4.015, 4.194) & (6.790, 7.216) & (8.710, 9.320) & (9.861, 10.654) & (10.429, 11.347) & (10.589, 11.560) \\ 
 	& 75 & (3.965, 4.157) & (6.635, 7.083) & (8.269, 8.968) & (9.041, 9.945) & (9.244, 10.222) & (9.277, 10.264) \\ 
 	& 80 & (3.805, 4.110) & (6.118, 6.809) & (7.202, 8.207) & (7.491, 8.631) & (7.528, 8.695) & (7.530, 8.698) \\ 
 	& 85 & (3.478, 3.958) & (5.078, 6.034) & (5.526, 6.666) & (5.576, 6.749) & (5.578, 6.752) &   \\ 
 	& 90 & (2.749, 3.416) & (3.487, 4.488) & (3.581, 4.643) & (3.583, 4.649) &  &  \\ 
 	& 95 & (1.998, 2.626) & (2.223, 3.015) & (2.232, 3.033) &  &  &   \\ 
 	& 100 & (1.268, 1.740) & (1.318, 1.821) &  &  &  &  \\ 
 	& 105 & (0.793, 1.108) &  &  &  &  &   \\ 
\\
  M 	& 60 & (4.174, 4.260) & (7.256, 7.463) & (9.504, 9.844) & (11.099, 11.579) & (12.144, 12.764) & (12.712, 13.468) \\ 
	& 65 & (4.126, 4.243) & (7.124, 7.406) & (9.240, 9.703) & (10.641, 11.287) & (11.396, 12.227) & (11.681, 12.612) \\ 
	& 70 & (4.059, 4.235) & (6.922, 7.342) & (8.773, 9.491) & (9.784, 10.777) & (10.170, 11.288) & (10.232, 11.374) \\ 
	& 75 & (3.913, 4.193) & (6.447, 7.128) & (7.848, 8.856) & (8.383, 9.551) & (8.477, 9.690) & (8.484, 9.702) \\ 
	& 80 & (3.633, 4.089) & (5.595, 6.577) & (6.279, 7.599) & (6.401, 7.807) & (6.408, 7.818) & (6.409, 7.819) \\ 
	& 85 & (3.061, 3.723) & (4.138, 5.280) & (4.333, 5.591) & (4.345, 5.617) & (4.345, 5.617) &  \\ 
	& 90 & (2.295, 2.965) & (2.715, 3.590) & (2.743, 3.632) & (2.744, 3.632) &  &  \\ 
	& 95 & (1.574, 2.033) & (1.664, 2.177) & (1.667, 2.181) &  &  &  \\ 
	& 100 & (0.920, 1.254) & (0.939, 1.281) &  &  &  &   \\ 
	& 105 & (0.563, 0.867) &  &  &  &  &   \\ 
\bottomrule
\end{tabular}
\end{small}
\end{table}

\end{appendices}

\newpage
\bibliographystyle{agsm}
\bibliography{cs_geometric.bib}

\end{document}